\documentclass[aps,prc,reprint,showpacs,groupedaddress,onecolumn]{revtex4-1}

\usepackage{graphicx,color} 
\usepackage{dcolumn}  
\usepackage{bm}       
\usepackage{amsmath}
\usepackage{xcolor}

\begin{document}

\newcommand {\nc} {\newcommand}

\newcommand{\vv}[1]{{$\bf {#1}$}}
\newcommand{\ul}[1]{\underline{#1}}
\newcommand{\vvm}[1]{{\bf {#1}}}
\def\btau{\mbox{\boldmath$\tau$}}

\nc {\IR} [1]{\textcolor{red}{#1}}
\nc {\IB} [1]{\textcolor{blue}{#1}}
\nc {\IP} [1]{\textcolor{violet}{#1}}
\nc {\IG} [1]{\textcolor{olive}{#1}}
\nc {\IT} [1]{\textcolor{teal}{#1}}

\title{Deuteron-alpha scattering: separable vs nonseparable Faddeev approach}

\author{L.~Hlophe$^{(a)}$}
\email{hlophe@nscl.msu.edu}
\author{Jin Lei$^{(b)}$}
\email{jinl@ohio.edu}
\author{Ch.~Elster$^{(b)}$}
\author{A. Nogga$^{(c)}$}
\author{F.M.~Nunes$^{(a)}$}
\author{D.~Jur\v{c}iukonis$^{(d)}$}
\author{A.~Deltuva$^{(d)}$}

\affiliation{
(a) National Superconducting Cyclotron Laboratory and Department of Physics and Astronomy, Michigan State University, East Lansing, MI 48824, USA \\
(b)Institute of Nuclear and Particle Physics,  and
Department of Physics and Astronomy,  Ohio University, Athens, OH 45701,
USA \\
(c) IAS-4, IKP-3, and JHCP,  Forschungszentrum J\"ulich, D-52428 J\"ulich, Germany\\
(d) Institute of Theoretical Physics and Astronomy, Vilnius University, Vilnius,
Lithuania\\
}

\date{\today}

\begin{abstract}
\begin{description}
\item[Background]
Deuteron induced reactions are widely used to probe nuclear structure
and astrophysical information.
Those (d,p) reactions may be viewed as three-body reactions and described with Faddeev techniques.

\item[Purpose] Faddeev-AGS equations in momentum space have a long tradition of
utilizing separable interactions in order to arrive at sets of coupled integral
equations in one variable. However, it needs to be demonstrated that observables calculated
based on separable interactions  agree exactly with
those based on nonseparable forces.

\item[Methods] Momentum space AGS equations are solved with separable and nonseparable forces as
coupled integral equations.

\item[Results] Deuteron-alpha scattering is calculated via momentum space AGS equations using the
CD-Bonn neutron-proton force and a Woods-Saxon type neutron(proton)-$^4$He force, for which the
Pauli-forbidden S-wave bound state is projected out. Elastic as well as breakup observables are
calculated and compared to results in which the interactions in the two-body sub-systems are
represented by separable interactions derived in the Ernst-Shakin-Thaler (EST) framework.

\item[Conclusions] We find that the calculations based on the separable representation of the
interactions and the original interactions give results that are in excellent agreement. Specifically, integrated cross sections and angular distributions for elastic scattering
agree within $\approx$ 1\%, which is well below typical experimental errors. In addition, the five-fold differential cross sections corresponding to breakup of the deuteron agree extremely well.


\end{description}
\end{abstract}

\pacs{24.10.-i,25.45.De,21.45.-v}

\maketitle

\section{Introduction}

Nuclear reactions offer an excellent probe into the properties of nuclei, particularly for
short-lived rare isotopes. Nuclear reactions are extremely proliferous:  they provide access
to structure properties that improve our knowledge of the underlying nucleon-nucleon (NN) force, they are used to populate excited states of interest, and through them we can extract astrophysical rates that cannot otherwise be directly measured.  However, a common concern when using nuclear reactions to extract nuclear properties pertains to the simplifications in describing the dynamics and consequent model dependence of the extracted properties. 
For this reason, it is a priority that our field develops new methods for reactions that are not
limited by unnecessary  approximations. 
In short, the goal is for a theory that includes  an exact treatment of the dynamics for the relevant degrees of freedom and
incorporates the  relevant reaction channels in a consistent framework and on an equal footing. In
addition, this reaction theory should be applicable across the nuclear chart and for a wide
range of energy regimes, so that it is not plagued by irreconcilable systematic 
differences~\cite{Atar:2018dhg}. 
This is the context of the present study. We focus first on deuteron induced
reactions on the $\alpha$-particle, but note that  the framework is readily extendable to heavier projectiles. 

While nuclear reactions are in a fundamental way many-body problems, in direct reactions, when
only a few degrees of freedom play a role, the problem is often reduced to a few-body problem.
As early as the pioneering work by Shanley~\cite{Shanley:1969zza},  three-body approaches  have
been successfully used to simplify the nuclear reaction problem and allow for an exact
treatment of the few-body dynamics. In Refs.~\cite{Shanley:1969zza,Miyagawa:1985js}, 
three-body solutions of the
Alt-Grassberger-Sandhas (AGS) equations~\cite{ags1967}  for the scattering of deuterons off an 
$\alpha$-particle were obtained using  rank-1 separable two-body forces. Both elastic
scattering observables and the total breakup cross sections, $n+p+\alpha$,  were computed and
compared to experiment. Despite its simplicity, the model provided a fair description of the data.

In the last decade, the AGS equations were  applied in the field of nuclear reactions by
Deltuva and Fonseca~\cite{deltuva:06b,deltuva2009a} without the need to employ separable
interactions, given the advances in computational power. The greatest challenge however,
when moving from applications in the few-nucleon sector to reactions with heavy ions concerns
the Coulomb force. Deltuva and Fonseca relied on screening the Coulomb interaction and subsequent renormalization to ensure
the compactness of the Faddeev equations. Over the last decade, there have been many
applications of this approach on a variety of reactions, such as deuteron stripping and
pickup reactions involving halo nuclei, with latest developments 
including also excitations of the nuclear core \cite{deltuva:17b}. Albeit the success of the method,
incorporating the Coulomb force still poses unwanted limitations: when it becomes too strong
(i.e. for heavier nuclei and for lower scattering energies), the Coulomb screening radius 
needed for an accurate description of the reaction increases to a point that renders the screening method ineffective. 
In practice this means that the methods in~\cite{deltuva:06b,deltuva2009a,deltuva:17b} have  been applied so far
to targets with mass $A \le 58$.  

An alternative to introducing Coulomb screening in the AGS equations is to cast the AGS
equations in the Coulomb basis.
In Ref.~\cite{akram2012},
Mukhamedzhanov and collaborators derived the AGS equations in the Coulomb basis,  indicating
also the necessity for employing a separable expansion of the forces in the different
subsystems. That work is focused on the formalism and includes no numerical applications. It thus remains to be proven that this alternative method indeed can provide precise solutions for deuteron induced reactions on heavy nuclei.

Our work represents an important stepping stone for using Faddeev techniques for nuclear
reactions with heavy nuclei. Starting from the non-relativistic AGS equations~\cite{ags1967},
we  make use of the formulation derived by Lovelace~\cite{lovelace1964},  and introduce
pair-wise separable interactions as previously
developed~\cite{hlophe2013,hlophe2014,hlophe2017}. For the treatment of the 
singularities in the
free three-body propagator above the breakup threshold, we implement a procedure proposed
in~\cite{Witala:2008my,Elster:2008hn}, which allows one to cast the so-called moving singularities
(see e.g.~\cite{Gloecklebook})
into singularities depending only on one variable, amenable to regular subtraction techniques.
In this work, we do not explicitly include target excitation in our formulation. However, 
the formalism readily extends to problems in which the two-body subsystems can be described as a coupled-channel problem involving the excitation of one of the bodies~\cite{hlophe2017est}.

In few-nucleon physics, the use of separable interactions is widespread and has been proven to
be accurate in describing neutron-deuteron processes below the pion
threshold~\cite{nemoto1998,Cornelius:1990zz}.
It is important that these benchmarks be performed for nuclear reactions, where the interaction
has larger complexity. Recently, we performed a benchmark for the bound state of $^{6}$Li~\cite{hlophe2017} and showed that one can obtain 4 digit accuracy in the binding energy if
including rank-8 potentials.  In the present work, we perform a similar benchmark but now for
deuteron-alpha elastic scattering and breakup reactions.  
We compare the results obtained in the new framework using separable interactions with the solutions obtained when no separable expansion is introduced~\cite{deltuva:06b}. 
The current work  includes only  nuclear interactions to allow for a careful inspection of the treatment of the short range aspects of the problem. We will focus on the inclusion of Coulomb in a subsequent study.

The paper is organized in the following way.
In Sec.~II, a brief summary of the theory is provided, highlighting essential ingredients for
our calculations. Section III introduces the two-body input to our calculations and presents
deuteron-alpha scattering observables for elastic as well as breakup scattering together with
essential convergence tests. We conclude in Sec. IV. A detailed description of our treatment
of the three-body breakup singularities and the calculation of the kinematical S-curve for
three particles with different masses is given in the appendices.

\section{Formal Considerations }

For the description of the scattering of a deuteron off an alpha particle, we employ the Alt,
Grassberger, and Sandhas (AGS) equations~\cite{AGS}, which are Faddeev-type integral equations in
momentum space for three-particle transition operators
\begin{equation}
 U^{ij} = \bar\delta_{ij}\;G_0^{-1} +\sum\limits_{k} \bar\delta_{ik} t_k G_0 \; U^{kj}.
\label{eq:2.1}
\end{equation}
Here $\bar\delta_{ij}=1-\delta_{ij}$ is the anti-Kronecker delta, $G_0(E)=(E+i0 -H_0)^{-1}$ is the free resolvent at the available three-particle center-of-mass energy $E$, and $H_0$ is the free three particle Hamiltonian. The three particles with masses $m_i$, $m_j$, and $m_k$ and spins $s_i$, $s_j$, and $s_k$ interact via pairwise forces $v^i \equiv v_{jk}$ ($i,j,k=1,2,3$ and cyclic permutations thereof) so that the operator $t_i(E) = v^i +v^i G_0(E) t_i(E)$ describes the two-body $t$~matrix in the subsystem $i$. The AGS transition amplitudes $U^{ij}$ are represented in their natural Jacobi coordinates $(p_i,q_i)$, where $p_i$ is the relative momentum of the $i^{th}$ pair and $q_i$ the momentum of the $i^{th}$ particle (also called the `spectator') relative to the pair. For solving the AGS equations, we choose a momentum space basis which depends on the magnitude of the momenta and angular momentum eigenstates. To proceed, we define the total spin $S_i=s_j+s_k$, relative orbital angular momentum $l_i$, and the total angular momentum by $J_i=l_i+S_i$ for the $i^{th}$ pair. 
The orbital angular momenta and spin of the spectator are represented by $\lambda_i$ and $s_i$ so
that the corresponding total angular momentum is ${\cal J}_i=\lambda_i+s_i$. The third component
of the spectator spin is denoted by $m_{s_i}$. The total angular momentum of the three-particle system is represented by $J$ and the third component by $M_J$. The states of conserved total angular momentum are thus given
as
\begin{equation}
|{p}_i{q}_i \alpha_{i} \rangle = |p_iq_i\;\big(l_i(s_is_j)S_i)J_{i} (\lambda_i s_i){\cal J}_i\big)\;JM_J
\rangle,
\label{eq:2.2}
\end{equation}
and are normalized as 
\begin{equation}
\langle {p}'_i{q}'_i \alpha_{i} ' | {p}_i{q}_i \alpha_{i} \rangle=
\frac{\delta(p'_i-p_i)}{p'_ip_i} \frac{\delta(q'_i-q_i)}{q'_iq_i} \delta_{\alpha_{i} '\alpha_{i}}.
\label{eq:2.3}
\end{equation}
The matrix elements of an AGS operator for the transition from an arrangement channel $j$ to $i$ is given by the expectation value $\langle \Phi_i|U^{ij}|\Phi_j \big\rangle$, where
\begin{eqnarray}
 |\Phi_i\rangle\equiv\sum\limits_{l_iS_i}\big|\phi_{{l_iS_i}}^{J_i}\;(l_iS_i)
J_{i}M_{J_i}\big\rangle|{\bf q}_i\;s_im_{s_i}\rangle,
\label{eq:2.4}
\end{eqnarray}
is the asymptotic state in arrangement channel $i$ with $|\phi_{l_iS_i}^{J_i}\big\rangle$ being the two-body bound state  wavefunction.   

Two approaches for solving Eqs.~(\ref{eq:2.1}) are adopted.
We shall refer to them as
the `separable' and the `non-separable' approach.  The former involves a separable
expansion of the two-body $t$~matrix while the latter does not. The non-separable method
is taken over from Ref.~\cite{deltuva:06b} and calculates fully off-shell $t$ matrices on the momentum grid, with subsequent interpolation using global spline functions whenever needed.

%
%
For the separable expansion method, the subsystem $t$~matrices take the well known form (see e.g.~\cite{Ernst:1973zzb}) 
\begin{equation}
t_{i}^{\alpha_{i} \alpha_{i}'} ( E_{q_i}) = \sum_{mn} |h^{\alpha_{i}}_m\rangle  
\tau_{mn}^{\alpha_{i} \alpha_{i}'}(E_{q_i})
\langle h^{\alpha_{i}'}_n| \ , 
\label{eq:2.5}
\end{equation}
where $|h^{\alpha_{i}}_m\rangle$ are the so-called form factors and $E_{q_i}\equiv E-q_i^2/2M_i$ represents the available two-body energy. Here $M_k$ is the reduced mass of the pair and the spectator. The indices ${m,n}$ represent the rank
of the separable potential, and $i$ stands for the arrangement channel. 
%
%
If the potential $v^{i}$ supports a bound state, the corresponding wave-function has the form
\begin{equation}
|\phi^i_{\alpha_i}\rangle=\sum\limits_{m}G_0(\epsilon_{\alpha_i})\;|h_{m}^{\alpha_i}\rangle\;c_{m\alpha_i},
\label{eq:2.6}
\end{equation}
where the two-body bound state energy is related to the on-shell
spectator momentum $q_{i0}$ by 
\begin{equation}
\epsilon_{\alpha_i} =E_{q_{i0}}= E -\frac{q_{i0}^2}{2M_i}. 
\label{eq:2.7}
\end{equation}
The constants $c_{m\alpha_i}$ are determined by substituting Eq.~(\ref{eq:2.6}) into a bound
state Lippmann-Schwinger equation. The partial wave matrix element for a transition from the bound state $|\phi^i_{\alpha_i}\rangle$ to $|\phi^j_{\alpha_j}\rangle$ becomes
\begin{equation}
\langle\phi^i_{\alpha_i}q_i\alpha_i|U^{ij}|\phi^j_{\alpha_j} q_j \alpha_j\rangle
=\sum\limits_{mn} c_{m\alpha_i}^*c_{n\alpha_j}
\langle h_m^{\alpha_i}\;q_i \alpha_i|G_0(\epsilon_{\alpha_i})U_{ij}
G_0(\epsilon_{\alpha_j})|h_n^{\alpha_j}q_j \alpha_j\rangle. 
\label{eq:2.8}
\end{equation}
To proceed, one defines the effective two-body AGS transition amplitudes~\cite{AGS}
%
%
\begin{equation}
 X_{m\alpha_i,n\alpha_j}^{ij}(q_i,q_j;E)\equiv 
\langle h_m^{\alpha_i}\;q_i \alpha_i|G_0(E)U^{ij}
G_0(E)|h_n^{\alpha_j}q_j \alpha_j\rangle ,
\label{eq:2.9}
\end{equation}
whose on-shell values coincide with those of Lovelace~\cite{Lovelace:1964} and
appear on the right-hand side of Eq.~(\ref{eq:2.8}). From Eqs.~(\ref{eq:2.1}), one obtains

\begin{eqnarray}
\langle h_m^{\alpha_i}q_i \alpha_i|G_0(E)U^{ij}G_0(E)|h_n^{\alpha_j}q_j\alpha_j\rangle &=&
\bar\delta_{ij} \;\langle h_m^{\alpha_i}q_i \alpha_i|G_0(E)|h_n^{\alpha_j}q_j\alpha_j\rangle\cr\cr
&+& \sum_k \sum_{\alpha_k\alpha_k'} \bar\delta_{ik} \int dq_k q_k^2\; 
 \langle h_m^{\alpha_i}q_i \alpha_i |G_0(E) | q_k \alpha_k\rangle\langle q_k \alpha_k|
t_{k}(E_{q_k})|q_k\alpha_k'\rangle \cr\cr
&& \times \langle q_k\alpha_k' | G_0(E) U^{kj}G_0(E)| h_n^{\alpha_j}q_i\alpha_j\rangle
\label{eq:2.10},
\end{eqnarray}
and substituting the separable two-body $t$~matrix yields

\begin{eqnarray}
\langle h_m^{\alpha_i}q_i \alpha_i|G_0(E)U^{ij}G_0(E)|h_n^{\alpha_j}q_j\alpha_j\rangle &=&
\bar\delta_{ij} \;\langle h_m^{\alpha_i}q_i \alpha_i|G_0(E)|h_n^{\alpha_j}q_j\alpha_j\rangle\cr\cr
&+& \sum_k \sum_{\alpha_k\alpha_k'}\sum_{m'n'} \bar\delta_{ik} \int dq_k q_k^2\; 
 \langle h_m^{\alpha_i}q_i \alpha_i |G_0(E) |h_{m'}^{\alpha_k} q_k \alpha_k\rangle\;
\tau^{\alpha_k\alpha_k'}_{m'n'}(E_{q_k})\; \cr\cr
&& \times \langle q_k\alpha_k' h_{n'}^{\alpha_k'} | G_0(E) U^{kj}G_0(E)| h_n^{\alpha_j}q_j\alpha_j\rangle.
\label{eq:2.11}
\end{eqnarray}
Defining the effective two-body `transition potentials'
\begin{align}
Z_{m\alpha_i,n\alpha_j}^{ij}(q_i,q_j,E) = \bar\delta_{ij} \;\langle h_m^{\alpha_i}q_i \alpha_i|G_0(E)|h_n^{\alpha_j}q_j\alpha_j\rangle,
\label{eq:2.12}
\end{align}
and using Eq.~(\ref{eq:2.9}), one can express Eqs.~(\ref{eq:2.11}) in a condensed form
\begin{align}
X_{m\alpha_i,n\alpha_j}^{ij}(q_i,q_j;E) & = Z_{m\alpha_i,n\alpha_j}^{ij}(q_i,q_j,E)
+ \sum_{k}\sum_{\alpha_k\alpha_{k}'}\sum_{m'n'} \int dq_k q_k^2\; Z_{m\alpha_i,m'\alpha_k}^{ik}(q_i,q_k;E) \nonumber \\
&\times \tau_{m'n'}^{\alpha_k\alpha_k'}(E_{q_k}) X_{n'\alpha_k',n\alpha_j}^{kj}(q_k,q_j;E).
\label{eq:2.13}
\end{align}
These equations are solved using iterative Lanzcos-type techniques~\cite{Saad:2003}.
The kernel contains two types of singularities, namely, the bound state pole and the
three-body breakup singularity. One one hand, the former constitutes a simple pole and is removed using standard  subtraction techniques. On the other hand, the three-body breakup pole has a complex structure
and its treatment is consequently more involved. A detailed discussion is provided in Appendix~\ref{sing}. 

 To evaluate the transition amplitude
for a breakup process, we first define the on-shell momentum of the pair $p_{i0}=\sqrt{2\mu_i(E-q_i^2/2M_i)}$, where $\mu_i$ is the reduced mass of the pair. The on-shell breakup amplitude is then expressed in terms of the amplitudes for elastic scattering and rearrangement processes using Eq.~(\ref{eq:2.1}) leading to
\begin{eqnarray}
\langle   q_i p_{i0} \alpha_i|U^{0j}|\Phi_j \big\rangle=\sum\limits_{k=1}^3 \langle  p_{0i}\; q_i\alpha_i|t_k\;G_0(E)\;U^{kj}|\Phi_j \big\rangle. 
\label{eq:2.14}
\end{eqnarray}
For separable two-body $t$ matrices, one obtains
\begin{eqnarray}
\langle   q_i p_{i0} \alpha_i|U^{0j}|\Phi_j \big\rangle=\sum\limits_{k=1}^3 \sum_{\alpha_k\alpha_k'\alpha_j}
\sum_{mn'n}\;\int dp_k p_k^2\;dq_k q_k^2\;\langle q_i p_{i0} \alpha_i|q_k p_k \alpha_k\rangle\; h_{m}^{\alpha_k}(p_k)\;\tau^{\alpha_k\alpha_k'}_{mn'}(E_{q_k})\;X_{n'\alpha_k',n\alpha_j}^{kj}(q_k,q_j;E)\;c_{n\alpha_j}.~~~~~~
\label{eq:2.15}
\end{eqnarray}
The matrix elements $\langle q_i p_{i0} \alpha_i|q_k p_k \alpha_k\rangle$ describe a transformation between two Jacobi coordinates for $i\ne k$ and are evaluated as described in
Ref.~\cite{Hlophe:2017bkd}. If $i=k$, these matrix elements reduce to the $\delta$-functions of Eq.~(\ref{eq:2.3}). It is thus seen that, once the effective two-body amplitudes $X^{ij}_{mn}$ have been determined, the transition amplitudes for elastic scattering, as well as rearrangement and breakup processes can be readily computed.

\section{Results and Discussion}
\label{discuss}

To demonstrate the accuracy of solving the Faddeev-AGS equations for deuteron-alpha scattering based on a 
separable expansion of the forces in the two-body subsystems, the convergence of the
expansion must be tested and finally the converged calculation must be compared to 
a numerically converged calculation based on the non-separable version of the same forces.
For this, we first define the three-body Hamiltonian with the forces in the different
two-body subsystems, namely the neutron-proton ($np$) force and the effective 
interaction between a neutron or
proton and the alpha-particle, i.e. the $n/p$-$\alpha$ forces.
Then, elastic as well as breakup observables for $d$-$\alpha$ scattering are calculated, and
their convergence with respect to the basis (rank) of the separable expansion is explored.
Finally, the well-converged separable calculations are compared to the corresponding ones
obtained with the non-separable forces in the subsystems.

\subsection{Two-body input}
\label{input}

For the $np$ force, the high precision CD-Bonn~\cite{Machleidt:2000ge} potential is adopted. 
As an effective interaction in the  $n/p$-$\alpha$ system, we employ the 
Bang potential as given in Ref.~\cite{Thompson2000} 
 consisting of an attractive  central Woods-Saxon  and spin-orbit terms. 
The two-body model space is restricted to $l_{i}\le 2$ for both the $np$ and $n/p$-$\alpha$ systems. The Coulomb repulsion in the $p$-$\alpha$ system is omitted. 
The Bang potential supports a bound state in the $n/p$-$\alpha$ S-wave channel,
which is removed with the projection technique described in~\cite{Hlophe:2017bkd}. 

For constructing  separable representations for the two-body $t$ matrices which enter the AGS
equations, we employ the method suggested by Ernst, Shakin, and Thaler~\cite{Ernst:1973zzb} (EST). 
The advantage of this scheme is 
that the  $t$ matrices calculated at specific energies with a given potential are taken as form factors of the
separable expansion. While in~\cite{Ernst:1973zzb} half-shell $t$ matrices are used, an extension to
using off-shell $t$ matrices is straightforward~\cite{Ernst:1974zza}. 
The details on the explicit construction of the separable representations employed here are laid out 
in~\cite{Hlophe:2013xca}. The application to the three-body calculation of the $^6$Li bound state is presented in~\cite{Hlophe:2017bkd}. For the convenience of the reader, as well as to establish notation, essential ingredients are briefly repeated.

The EST approach applied in momentum space requires solving a two-body Lippmann-Schwinger (LS) equation 
at a specific scattering energy $E_n$ with a given
real (or complex) potential, leading to a $t$ matrix $t(p,p_n;E_n)$. The on-shell $t$ matrix corresponds to momentum $p_n=\sqrt{2\mu E_n}$. For all other values of $p_n$, the $t$ matrix is fully off-shell. We use  $E_n$ (support energies) and $p_n$  (support momenta) independently to construct the separable interactions.  
Thus any specific solution of the LS equation is characterized by the pair ($E_n$, $p_n$), which we call
an EST support point. Naturally, when $E_n$ corresponds to a bound state energy in the two-body system, the
$t$ matrix is always fully off-shell. 
In Ref.~\cite{Hlophe:2017bkd}, we calculated the three-body binding energy of $^6$Li  and found that by
choosing $p_n$ independently from the bound state energy, we could achieve better
accuracy.

To ensure that the separable expansion converges, it is imperative to calculate observables using
successively increasing ranks. For this reason, we define separable representations of the CD-Bonn and
Bang potentials with ranks ranging from 3 to 7. Table~\ref{table1} lists the EST support points used to
construct the separable representations of the CD-Bonn potential, while  Table~\ref{table2} shows those
for the Bang potential. The separable representations provide a good description of the two-body $t$ matrix over a range of relative two-body 
energies $E_{2}$ which, in the context of solving the Faddeev-AGS equations, corresponds to $-\infty < E_{2} \le E$  ($E$ being the three-body energy in the c.m. frame).

\subsection{Deuteron-alpha scattering observables}

The validity of the separable expansion depends on the beam energy for the reactions, thus
we choose three deuteron beam energies, $E_d=10,\;20,$ and$\;50$~MeV, for the benchmark 
calculations of the separable vs. non-separable solutions of the AGS equations. These energies cover the typical range for experiments of (d,p) reactions.

\subsubsection{Integrated cross sections}
\label{obsv:intg}

Integrated cross sections are an important test of the calibration of our methods.  We aim to achieve a precision of $\approx 1$\% on this observables which is well below the typical experimental errors and the uncertainties associated with the two-body interactions. Tables~\ref{table3} and \ref{table4} show the integrated cross sections computed using the separable potentials given in Tables~\ref{table1} and~\ref{table2} at 10 and 20~MeV incident
deuteron energy. We show results for separable potentials constructed only with support momenta obeying the constraint $p_n=\sqrt{2\mu|E_n|}$ (NN-EST3-1, NN-EST4-1, NN-EST5-1, NN-EST6-1, NN-EST7-1, NA-EST3-1, NA-EST4-1, NA-EST5-1, NA-EST6-1, and NA-EST7-1) as in~\cite{Cornelius:1990zz}. We also show results for separable potentials for which the support momenta $p_n$ are independent from $E_n$ (NN-EST6-2, NN-EST6-3, NN-EST7-2, NA-EST6-2, NA-EST6-3,
and NA-EST7-2). These results are benchmarked against  the integrated cross sections obtained using the non-separable approach, given in
the bottom row. The first point to make is that a similar convergence rate is observed for both elastic and breakup cross sections. Moreover, the convergence pattern is similar for both the
10 and 20~MeV deuteron beam energies.
The second point worth making is that, in this case, the inclusion of off-shell momenta does not represent an improvement of the restricted $p_n=\sqrt{2\mu|E_n|}$ basis, contrary to our observation for bound states. Thirdly, a rank-6 interaction already provides the desired level of precision.
Finally, and most importantly, the results agree with the cross sections obtained from the non-separable calculation  to $\approx 1$\%.

 It is interesting to contrast our findings with those
 of Ref.~\cite{Cornelius:1990zz} for $nd$ scattering at 10~MeV incident neutron energy. In that work the authors demonstrated
 that an EST rank-3 potential could reproduce the integrated cross sections calculated
 using the non-separable method to about 2~$\%$. From Tables~\ref{table3} and~\ref{table4},
 we see that the discrepancy between results calculated with the separable and non-separable
approach is already better than $\sim 1.5\%$ for the rank-4 potentials, indicating that our findings are
consistent with those of Ref.~\cite{Cornelius:1990zz}.

\subsubsection{Elastic scattering: deuteron angular distributions}
\label{obsv:el}

\begin{figure}
\centering
\includegraphics[width=12cm]{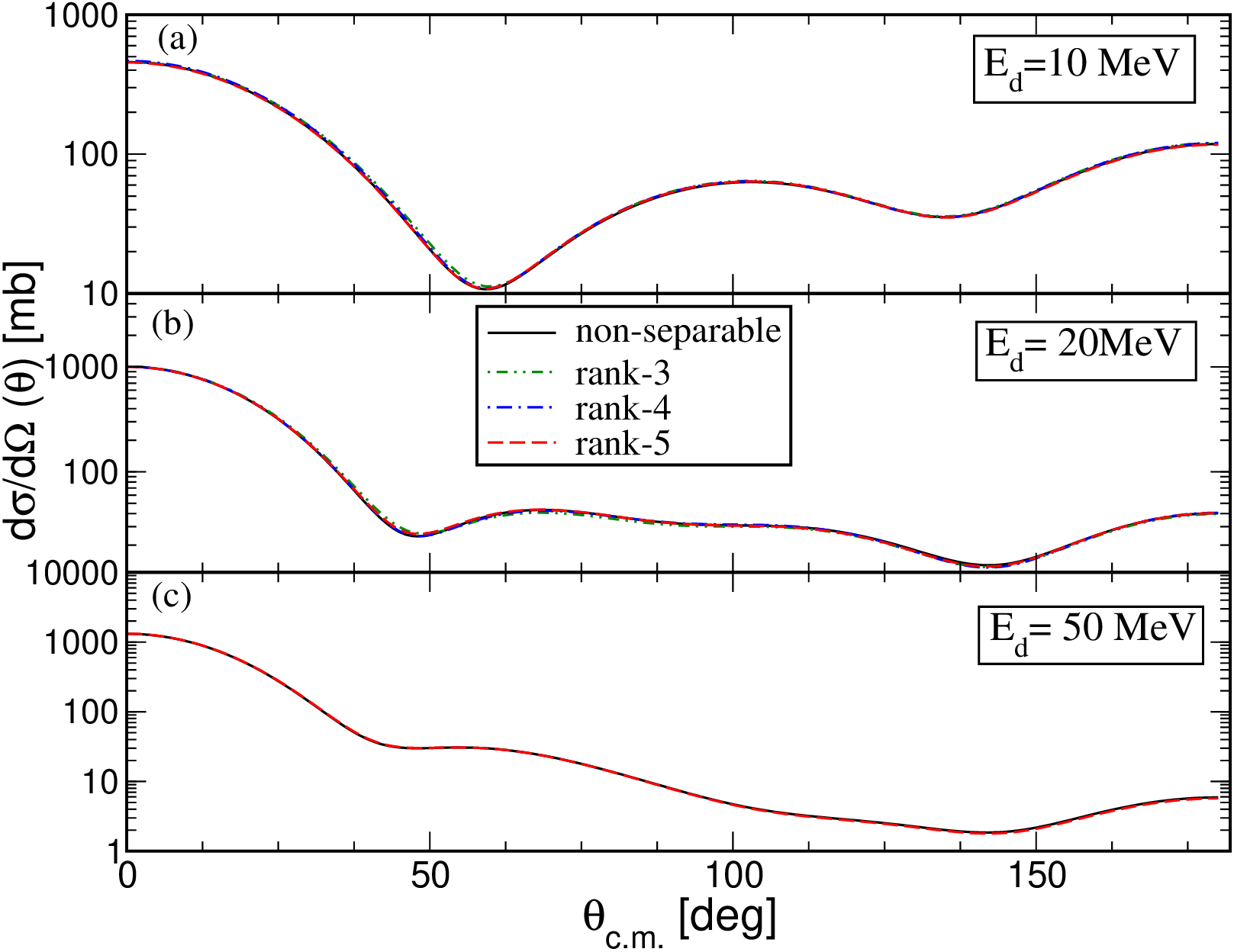}
\caption{The differential cross section for elastic deuteron-alpha scattering as a function of the center-of-mass angle $\theta_{c.m.}$. 
Panels (a) and (b) display results for the incident deuteron energies $E_d=$~10 and 20~MeV as indicated in the figure. The solid curve shows cross sections computed using the non-separable approach. The results calculated with the rank-3, rank-4, and rank-5 potentials are illustrated by the dot-dot-dashed, dot-dashed, and dashed curves. Panel (c) depicts the differential cross section at $E_d=$~50~MeV.
The solid and dashed curves indicate the converged results evaluated via the separable and non-separable
approach.
%
}.
\label{fig1}
\end{figure}

Next, we consider the differential cross sections for elastic deuteron-alpha
scattering for three deuteron beam energies $E_d$. 
In Fig.~\ref{fig1}, the differential cross sections for elastic $d+\alpha$ scattering as a function of
the center-of-mass (c.m.) angle $\theta_{c.m.}$ are shown for 
three different incident deuteron energies, $E_d=$~10, 20~MeV, and 50~MeV. The solid lines indicate the
angular distributions evaluated using the non-separable approach. The calculations
 obtained using rank-3, rank-4, and rank-5 separable potentials are indicated by the dash-dot-dotted,
dash-dotted, and dashed lines. All four lines are nearly indistinguishable which demonstrates that the
separable expansion is not only well-converged, but also  yields deuteron angular distributions that are in excellent agreement with those obtained using the non-separable approach. The relative differences between the two results remain at $\approx 1\;$\% throughout the angular
range. This is well below the usual experimental uncertainties for this observable.   
It is worth noting that the separable expansion
method is not limited to the low beam energies mentioned above. Increasing the beam energy by tens of
MeV does not introduce any principal technical difficulties except that, obviously, the dependence on the rank of the
separable expansion needs to be reevaluated.

\subsubsection{Deuteron breakup: fivefold differential cross section}
\label{obsv:bu}

 For an even more stringent test of the separable expansion method, we explore the
 convergence of the five-fold breakup differential cross section with respect to the rank
 of the separable expansion. Typically, one proceeds by specifying configurations
 defined by the outgoing proton and alpha particle angles $(\theta_p,\;\phi_p)$ and  $(\theta_\alpha,\;\phi_\alpha)$. The corresponding S-curves are then evaluated
 using energy and momentum conservation as described in Appendix~\ref{scurve}.
 Each configuration is given in the format  $(\theta_\alpha,\;\phi_\alpha;\;\theta_p,\;\phi_p)$, with the angles given in degrees. We consider two configurations, one of which corresponds to the final state interaction (FSI). The FSI configurations are defined such that
$E_{np}\approx 0$ for a single value of the arclength $S$.
  Here $E_{np}$ is the relative energy between the outgoing neutron and proton.  

\begin{figure}
\begin{center}
\includegraphics[width=12cm]{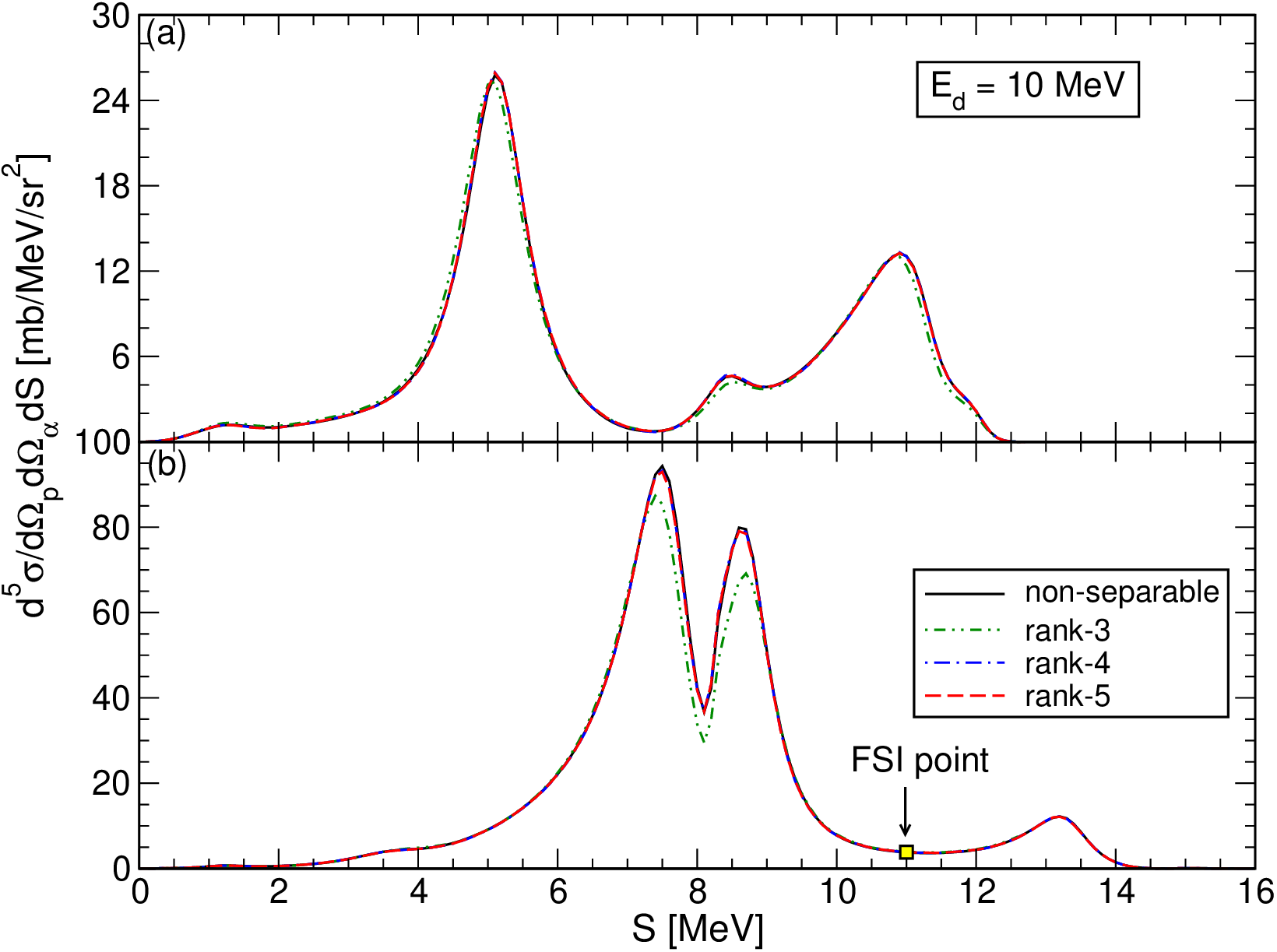}
\vspace{3mm}
\caption{The five-fold differential cross section for 
deuteron-alpha scattering as a function of the arc-length $S$ at 10~MeV
incident deuteron energy. Panel (a) shows results for the configuration $(25.6^{\circ},\;0^{\circ};\;63.6^{\circ},\;180^{\circ})$ while panel (b) depicts results for the FSI configuration $(31.4^{\circ},\;0^{\circ};\;5.1^{\circ},\;180^{\circ})$. 
The solid line corresponds to results calculated using the non-separable approach
while the ones computed via the separable expansion method are illustrated by the dash-dot-dotted, dash-dotted, and dashed lines for the rank-3, rank-4, and rank-5 separable potentials, respectively. The QFS
and FSI points are indicated by the symbols. 
}
\label{fig2}
\end{center}
\end{figure}

Figure~\ref{fig2} displays the five-fold breakup differential cross section on the S-curve for
$E_d=10$~MeV for two different configurations of the $\alpha+n+p$ system. Panel (a) shows results for the configuration $(25.6^{\circ},\;0^{\circ};\;63.6^{\circ},\;180^{\circ})$ while panel (b) depicts results for the FSI configuration $(31.4^{\circ},\;0^{\circ};\;5.1^{\circ},\;180^{\circ})$.
The cross sections computed with the non-separable method are indicated by the
solid line. Results evaluated using the separable expansion method are depicted
by the dash-dot-dotted, dash-dotted, and dashed lines for the rank-3, rank-4, and rank-5
potentials of Tables~\ref{table1} and \ref{table2}. The FSI point is indicated by the filled square.
The deviation of the dash-dot-dotted line from the other curves in both panels
demonstrates that the rank-3 potential is inadequate for this observable. However, the separable expansion is still rapidly converging so that the rank-4 and rank-5 curves are virtually indistinguishable. We also observe that the converged results from the separable approach are in excellent agreement with those calculated using the non-separable method.   
    
\begin{figure}
\begin{center}
\includegraphics[width=12cm]{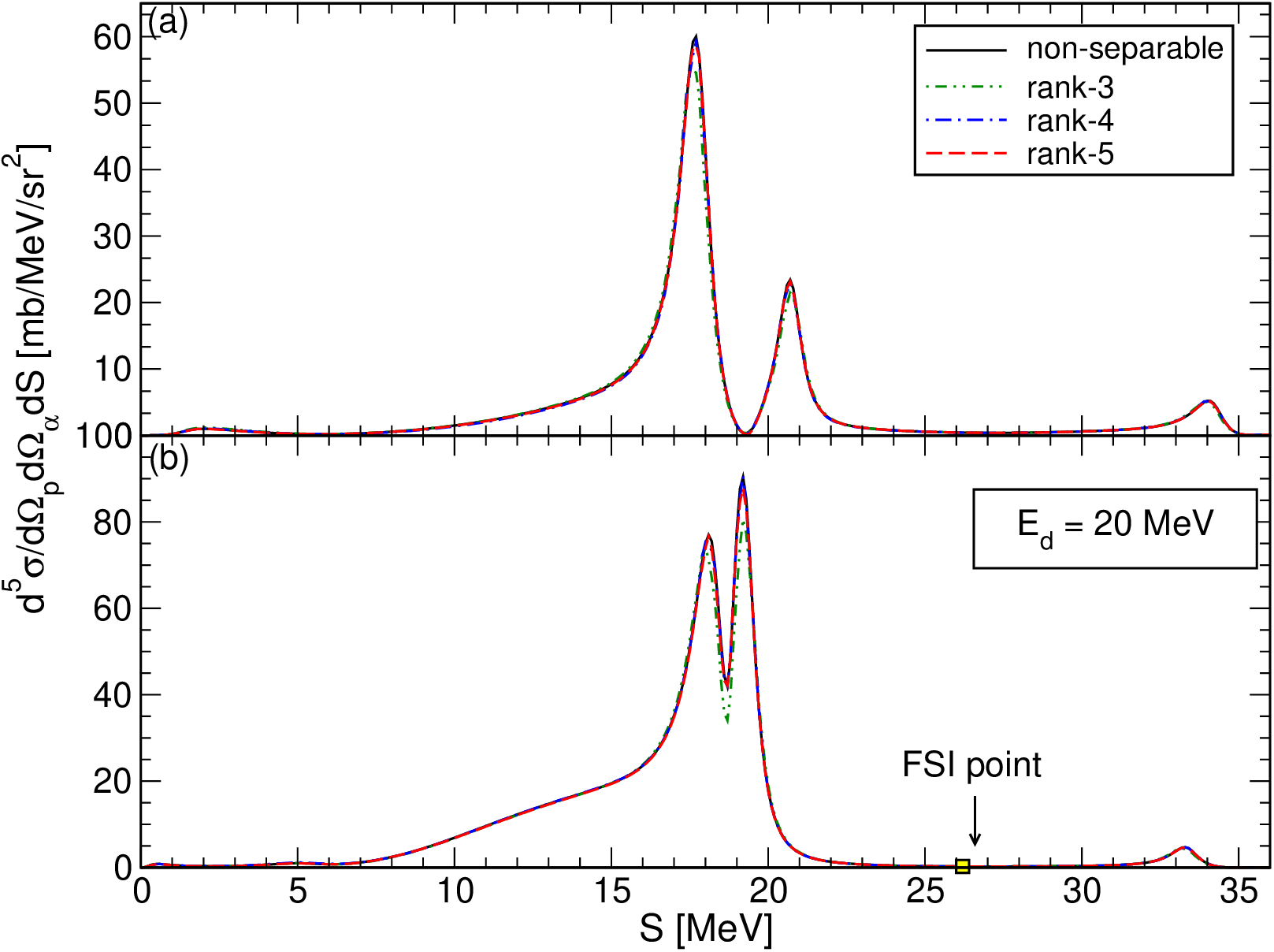}
\vspace{3mm}
\caption{Same as Fig,~\ref{fig2} but for 20~MeV incident deuteron energy. Panel (a) shows results for the configuration $(29.0^{\circ},\;0^{\circ};\;22.5^{\circ},\;180^{\circ})$ while panel (b) depicts results for the FSI configuration  and $(25.6^{\circ},\;0^{\circ};\;1.7^{\circ},\;180^{\circ})$.
}
\label{fig3}
\end{center}
\end{figure}

Next, we consider the five-fold breakup differential cross section at 20~MeV incident deuteron energy. Figure~\ref{fig3} is the same as Figure~\ref{fig2} but for
$E_d=20$~MeV. The configurations depicted in panels (a) and (b) are $(29.0^{\circ},\;0^{\circ};\;22.5^{\circ},\;180^{\circ})$ and $(25.6^{\circ},\;0^{\circ};\;1.7^{\circ},\;180^{\circ})$, where
the latter corresponds to FSI. We note that the convergence pattern
is similar to that of $E_d=10$~MeV. Also, the rank-4 and rank-5 results are in very good agreement with those obtained via the non-separable approach.

In Section~\ref{obsv:el} we showed that the agreement between deuteron angular distributions calculated using the separable and non-separable methods remains
excellent when the beam energy is increased to $E_d=$~50~MeV. For completeness,
it is imperative that we also compare the five-fold breakup differential cross sections at this energy.   

\begin{figure}
\begin{center}
\includegraphics[width=15cm]{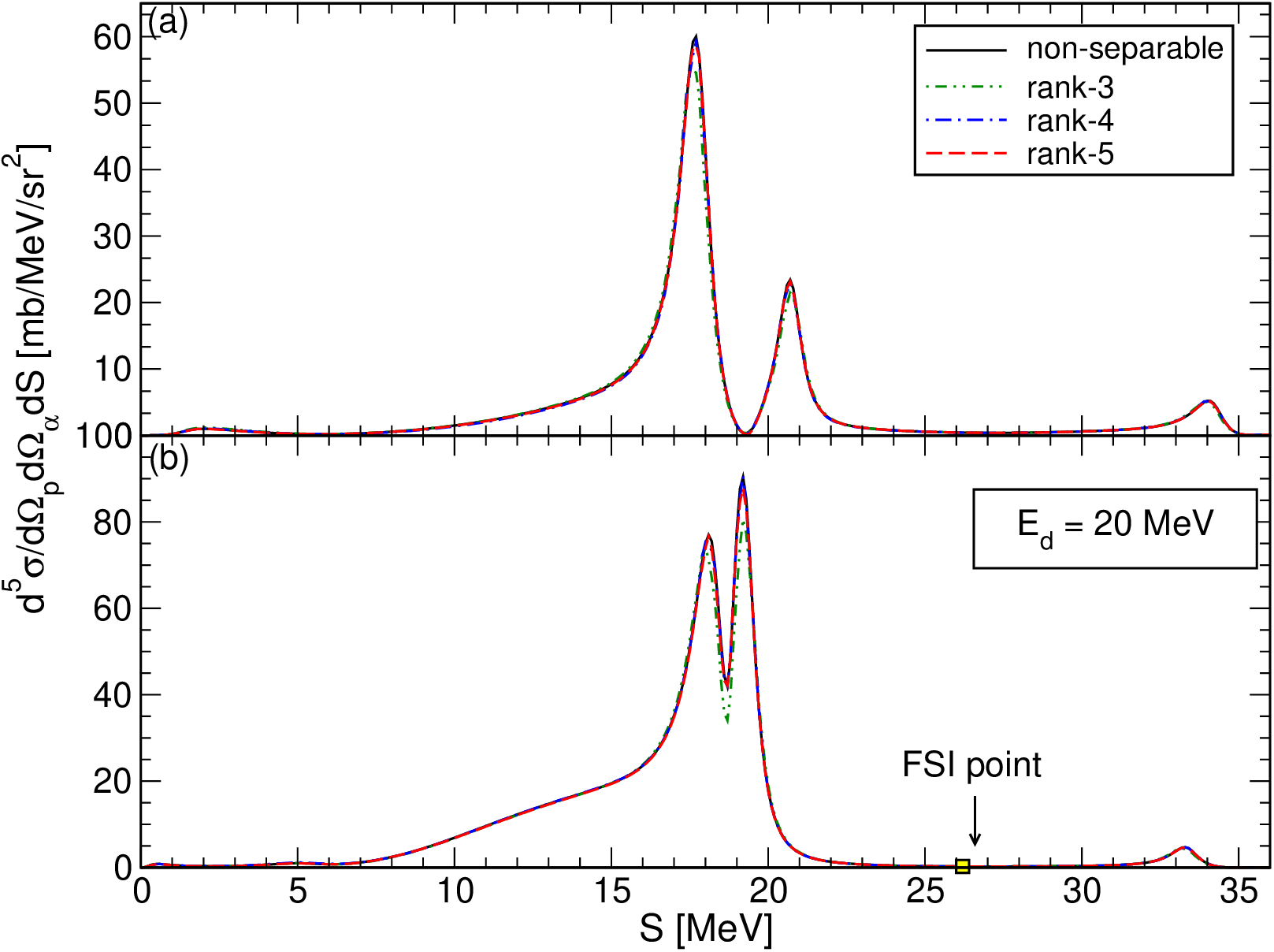}
\vspace{3mm}
\caption{The five-fold differential cross section for 
deuteron-alpha scattering as a function of the arc-length $S$ at 50~MeV
incident deuteron energy. Panel (a) depicts the configuration $(14.0^{\circ},\;0^{\circ};\;7.2^{\circ},\;180^{\circ})$
while the off-plane configurations $(14.0^{\circ},\;0^{\circ};\;7.2^{\circ},\;120^{\circ})$ and
$(22.2^{\circ},\;0^{\circ};\;104.4^{\circ},\;100^{\circ})$
are shown in panels (b) and (c). The FSI configuration $(22.2^{\circ},\;0^{\circ};\;104.4^{\circ},\;180^{\circ})$  is illustrated in panel (d) and the FSI point is indicated by the filled square. The solid line corresponds to results calculated using the non-separable approach while the dashed lines depicts those computed via the separable expansion method.  
}
\label{fig4}
\end{center}
\end{figure}

The comparison is shown in Fig.~\ref{fig4}.
 Panel (a) depicts the configuration $(14.0^{\circ},\;0^{\circ};\;7.2^{\circ},\;180^{\circ})$
while the off-plane configurations $(14.0^{\circ},\;0^{\circ};\;7.2^{\circ},\;120^{\circ})$ and
$(22.2^{\circ},\;0^{\circ};\;104.4^{\circ},\;100^{\circ})$
are shown in panels (b) and (c). The FSI configuration $(22.2^{\circ},\;0^{\circ};\;104.4^{\circ},\;180^{\circ})$  is illustrated in panel (d). The filled square indicates the FSI point. The solid line corresponds to results calculated using the non-separable approach while the dashed lines depicts those computed via the separable expansion method. We observe that the two methods are in excellent agreement and that the level of agreement is consistent with the one obtained for 
$E_d=$~10 and 20~MeV.

\section{Summary and Outlook}

Deuteron induced nuclear reactions are a widely used probe in nuclear physics. From a theoretical standpoint, these reactions are often mapped on a three-body problem $n+p+A$. The exact solution of the three-body problem can be obtained using the momentum-space Faddeev AGS framework.
Given the long-range Coulomb force, the standard implementation of the AGS equations
for deuteron induced nuclear reactions relies on the screening and renormalization
method. However this limits the application to lighter targets and/or higher beam
energies. To circumvent this limitation, one can instead use separable interactions
and the Coulomb distorted basis as proposed in \cite{akram2012}. 

In this study, we implemented an AGS framework based on EST-like  separable interactions, solved those separable AGS equations for scattering and constructed elastic and breakup observables.  For the purpose of this benchmark, no Coulomb interactions were included. We applied our method to $d-\alpha$ scattering for $E_d=10,\; 20,\; 50$ MeV, taking the same interactions as those used in a previous work where we benchmarked the use of separable interactions for three-body bound states \cite{Hlophe:2017bkd}. We find that the new method converges well. Depending on the desired observable, rank-4 or rank-6 were sufficient for obtaining 1\% precision.  

For the benchmark, we also performed the AGS calculations  using the non-separable interactions.  The results obtained with the separable interactions agree very well with those using the standard non-separable AGS method. Specifically, the total cross sections and the elastic differential angular distributions calculated using these two methods agree to $1$\%.  Breakup is typically harder to calculate precisely.
For this reason,  we selected a wide variety of configurations for proton and neutron angles in order to test the method in the most extreme limits. For all cases considered, we found that the five-fold differential breakup cross sections obtained using the separable framework for rank-4 was already in very good agreement with the results using the 
non-separable approach.  Moreover, no special adjustment of the separable force was needed when going from the three-body bound state calculation to the scattering application.

Consistent with the conclusion from the study of $^6$Li bound states \cite{Hlophe:2017bkd}, we here demonstrate that the separable formulation  provides a reliable method to solve the three-body scattering problem. It is now possible to follow onto the final step of the process, namely the inclusion of the Coulomb potential by solving the Faddeev-AGS equations in the Coulomb basis.

\appendix
\section{Treatment of Three-Body Breakup Singularities}

\label{sing}
The kernel of Eqs.~(\ref{eq:2.13}) contains bound state singularities as well as three-body breakup poles. The former are simple poles and can be removed from the kernel using standard subtraction procedures. The breakup singularity has a more complicated structure
and its removal requires more work. To proceed, we first note that the matrix elements of the three-body propagator have the form
\begin{align}
 G_0(p_i,q_i;E)=\Big[ E-\frac{p_i^2}{2\mu_i}-\frac{q_i^2}{2M_i}+i0\Big]^{-1},
\label{eq:sing1}
\end{align}
and that the breakup pole is located at the on-shell pair momentum 
$p_{0i}^2(q_i)=2\mu_i(E-q_i^2/2M_i)$, so that Eq.~(\ref{eq:sing1}) can
be written as
\begin{align}
G_0(p_i,q_i;E)=\frac{2\mu_i}{p_{0i}^2(q_i)-p_i^2+i0}.
\label{eq:sing2}
\end{align}
The breakup pole constitutes a `moving singularity' due to its dependence on the spectator momentum and is the major cause of numerical difficulties for scattering energies above the three-body breakup threshold. The numerical
complications are manifest when attempting to compute the transition potentials 
via Eq.~(\ref{eq:2.12}). Using the fact that
\begin{align}
 \langle p_i q_i\alpha_i|p_k q_k\alpha_k\rangle= \int_{-1}^{1}dx\; 
  G_{\alpha_i\alpha_k}(q_i,q_k,x)\frac{\delta(p_i-\pi_i(q_i,q_k,x))}{p_i^2}\frac{\delta(p_k-\pi_k(q_i,q_k,x))}{p_k^2},
\label{eq:sing3}
\end{align}
Eq.~(\ref{eq:2.12}) can be expressed as
\begin{align}
Z_{m\alpha_i,n\alpha_k}^{ik}(q_i,q_k,E)= \int_{-1}^{1}dx\;   h_m^{\alpha_i} (\pi_i)
G_{\alpha_i\alpha_k}(q_i,q_k,x) G_0\big(\pi_i,q_i;E\big) h_n^{\alpha_k}(\pi_k),
\label{eq:sing4}
\end{align}
where the shifted momenta $\pi_i=\sqrt{\beta^2 q_i^2+q_k^2+2\beta  q_i  q_k x}$ 
and $\pi_k=\sqrt{q_i^2+\eta^2 q_k^2+2\eta  q_i  q_k x}$. Here  $\beta=\mu_i/m_k$ and $\eta=\mu_j/m_k$. Substituting
Eq.~(\ref{eq:sing2}) into Eq.~(\ref{eq:sing4}) yields
\begin{align}
Z_{m\alpha_i,n\alpha_k}^{ik}(q_i,q_k,E)= \int_{-1}^{1}dx\;   
h_m^{\alpha_i} (\pi_i)\;G_{\alpha_i\alpha_k}(q_i,q_k,x) 
\frac{2\mu_i}{p_{0i}^2-\pi_i^2(q_i,q_k,x)+i0}\; h_n^{\alpha_k}(\pi_k).
\label{eq:sing5}
\end{align}
Defining  
\begin{align}
x_0=\frac{p_{0i}^2-\beta^2 q_i^2-q_k^2}{2\beta q_i q_k},
\label{eq:sing6}
\end{align}
Eq.~(\ref{eq:sing5}) can be expressed in the form
\begin{align}
Z_{m\alpha_i,n\alpha_k}^{ik}(q_i,q_k,E)= \int_{-1}^{1}dx\;  
h_m^{\alpha_i} (\pi_i) G_{\alpha_i\alpha_k}(q_i,q_k,x)
\frac{\mu_i}{\beta q_iq_k}\frac{1}{x_0-x+i0}  h_n^{\alpha_k}(\pi_k).
\label{eq:sing7}
\end{align}
The presence of $1/(x_0-x+i0)$ implies a logarithmic singularity
which occurs only if $q_i$ is
below $q_{bi}=\sqrt{2M_i E}$ so that the on-shell momentum of the pair 
$p_{0i}(q_i) > 0$. The transition potentials are thus well-defined for $q_i> q_{bi}$ and can be computed in the usual manner according to Eq.~(\ref{eq:sing4}).
For  $q_i \le q_{bi}$ the $x$-integration is singular and an appropriate regularization scheme can be applied for $|x_0| < 1$. Contrarily, the Cauchy principal value is undefined for $|x_0|=1$ due to the presence of end point singularities. As a way forward,
we substitute Eqs.~(\ref{eq:2.12}) into the kernel of Eqs.~(\ref{eq:2.13}) leading to
\begin{align}
X_{m\alpha_i,n\alpha_k}^{ik}(q_i,q_k;z)&= Z_{m\alpha_i,n\alpha_k}^{ik}(q_i,q_k,E)
+\sum_{j}\sum_{\alpha_j\alpha_j'}\sum_{m'n'}\bar\delta_{ij}
\int dp_i p_i^2 \;dq_j q_j^2\; dp_j p_j^2\;   h_m^{\alpha_i}(p_i)
\frac{2\mu_j}{p_{0j}^2(q_j)-p_j^2+i0} \nonumber \\
& \times  \langle q_ip_i \alpha_i|q_jp_j\alpha_j\rangle
h_{m'}^{\alpha_j}(p_j)
\tau_{m'n'}^{\alpha_j\alpha_j'}(E_{q_j})X_{m'\alpha_j',n\alpha_k}^{jk}(q_j,q_k;E).
\label{eq:sing8} 
\end{align}
For the partial waves containing at least one two-body bound state, the coupling matrix takes the form
\begin{align} 
\tau_{m'n'}^{\alpha_j\alpha_j'}(E_{q_j})\equiv\frac{2M_j}{q_{0j}^2-q_j^2+i0}\tilde\tau_{m'n'}^{\alpha_j\alpha_j'}(E_{q_j}),
\label{eq:sing9} 
\end{align}
so that
\begin{align}
X_{m\alpha_i,n\alpha_k}^{ik}(q_i,q_k;z)&= Z_{m\alpha_i,n\alpha_k}^{ik}(q_i,q_k,E)
+\sum_{j}\sum_{\alpha_j\alpha_j'}\sum_{m'n'}\bar\delta_{ij}
\int dp_i p_i^2 \;dq_j q_j^2\; dp_j p_j^2 
;  h_m^{\alpha_i}(p_i)
\frac{2\mu_j}{p_{0j}^2(q_j)-p_j^2+i0} \nonumber \\
& \times  \langle q_ip_i \alpha_i|q_jp_j\alpha_j\rangle\;
h_{m'}^{\alpha_j}(p_j)
\frac{2M_j}{q_{0j}^2-q_j^2+i0}\tilde\tau_{m'n'}^{\alpha_j\alpha_j'}(E_{q_j})
X_{m'\alpha_j',n\alpha_k}^{jk}(q_j,q_k;E).
\label{eq:sing10} 
\end{align}
Furthermore, the coupling matrix for partial waves supporting more than one bound state can always be written as a sum of terms similar to the right-hand side of Eq.~(\ref{eq:sing9}) by utilizing the concept of partial fractions~\cite{Witala:2008my,Elster:2008hn}. The most general singularity structure of the kernel is of the form
\begin{align}
\frac{2\mu_j}{p_{0j}^2(q_j)-p_j^2+i0}\frac{2M_j}{q_{0j}^2-q_j^2+i0}.
\label{eq:sing11} 
\end{align}
The bound state pole is disentangled from the propagator singularity by 
using the partial fraction expansion
\begin{align}
\frac{2\mu_j}{p_{0j}^2(q_j)-p_j^2+i0}\frac{2M_j}{q_{0j}^2-q_j^2+i0}
=\frac{1}{\frac{{p_j}^2}{2\mu_j}-\epsilon_{\alpha_j}}\frac{2\mu_j}{p_{0j}^2(q_j)-p_j^2+i0}-
 \frac{1}{\frac{{p_j}^2}{2\mu_j}-\epsilon_{\alpha_j}}\frac{2M_j}{q_{0j}^2-q_j^2+i0}. 
\label{eq:sing12} 
\end{align}
Since the factor $1/(\frac{{p_j}^2}{2\mu_j}-\epsilon_{\alpha_j})$ is non-singular, the
desired separation of the singularities is achieved. Substituting Eq.~(\ref{eq:sing12}) into Eq.~(\ref{eq:sing10}) and using Eq.~(\ref{eq:sing3}) one obtains
\begin{align}
X_{m\alpha_i,n\alpha_k}^{ik}(q_i,q_k;E)& = Z_{m\alpha_i,n\alpha_k}^{ik}(q_i,q_k,E)+
\sum_j \sum_{\alpha_j\alpha_j'}\sum_{m'n'} \bar\delta_{ij} \Bigg\{
 \nonumber \\
&\int dp_i p_i^2\; dq_j q_j^2\; dp_j p_j^2\; \int\limits_{-1}^{1}dx\; h_m^{\alpha_i}(p_i)
\frac{1}{\frac{{p_j}^2}{2\mu_j}-\epsilon_{\alpha_j}}  \frac{2\mu_j}{p_{0j}^2(q_j)-p_j^2+i0} h_{m'}^{\alpha_j}(p_j)
G_{\alpha_i\alpha_j}(q_i,q_j,x) \nonumber \\
&\times \frac{\delta\big(p_i-\pi_i(q_i,q_j,x)\big)}{p_i^2}
\frac{\delta\big(p_j-\pi_j(q_i,q_j,x)\big)}{p_j^2} \tilde\tau_{m'n'}^{\alpha_j\alpha_j'}(E_{q_j})
X_{n'\alpha_j',n\alpha_k}^{jk}(q_j,q_k;E)\nonumber \\
& - \int dp_i  p_i^2\;  dq_j  q_j^2\; dp_j  p_j^2\;  \int\limits_{-1}^{1}dx\;  h_m^{\alpha_i}(p_i)
\frac{1}{\frac{{p_j}^2}{2\mu_j}-\epsilon_{\alpha_j}}\frac{2M_j}{q_{0j}^2-q_j^2+i0}
h_{m'}^{\alpha_j}(p_j)  G_{\alpha_i\alpha_j}(q_i,q_j,x) \nonumber \\
&\times \frac{\delta\big(p_i-\pi_i(q_i,q_j,x)\big)}{p_i^2} \frac{\delta\big(p_j-\pi_j(q_i,q_j,x)\big)}{p_j^2}
\tilde\tau_{m'n'}^{\alpha_j\alpha_j'}(E_{q_j})X_{n'\alpha_j',n\alpha_k}^{jk}(q_j,q_k;E) \Bigg\}.
\label{eq:sing13}
\end{align}
For the first term, we use the delta function to evaluate the $p_j$ integral 
and convert the delta function in $p_i$ to a delta function in $x$
\begin{align}
\frac{\delta(p_i-\pi_i(q_i,q_j,x))}{p_i^2}=\frac{1}{\beta  p_i q_i q_j}\delta(x-x_0).
\label{eq:sing14}
\end{align}
For the second term inside the curly brackets, we use the delta functions 
to evaluate the $p_i, p_j$ integrals and perform
the angular integration to obtain different transition potentials
\begin{align}
\bar Z_{m\alpha_i,m'\alpha_j}^{ij}(q_i,q_j,E)\equiv \int\limits_{-1}^{1}dx\;  h_m^{\alpha_i}(\pi_i) 
 \frac{1}{\epsilon_{\alpha_j}+\frac{{\pi_j}^2}{2\mu_j}} h_{m'}^{\alpha_j}(\pi_j)\;G_{\alpha_i,\alpha_j}(q_i,q_j,x).
\label{eq:sing15}
\end{align}
Substituting Eqs.~(\ref{eq:sing14}) and~(\ref{eq:sing15}) into
Eq.~(\ref{eq:sing13}), and using the fact that kinetic energy is the same in all three Jacobi coordinates leads to
\begin{align}
X_{m\alpha_i,n\alpha_k}^{ik}(q_i,q_k;E)& = Z_{m\alpha_i,n\alpha_k}^{ik}(q_i,q_k,E)+
\sum\limits_{j} \sum\limits_{\alpha_j\alpha_j'} \sum\limits_{m'n'} \bar\delta_{ij}\Bigg\{ \nonumber \\
& \int dp_i  p_i\;   \frac{1}{\beta q_i}  h_m^{\alpha_i}(p_i)
 \frac{2\mu_i}{p_{0i}^2(q_i)-p_i^2+i0} \int\limits_{q_j=|p_i-\beta q_i|}^{q_j=p_i+\beta q_i} dq_j q_{j} \frac{1}{\frac{{p_j}^2}{2\mu_j}-\epsilon_{\alpha_j}}
h_{m'}^{\alpha_j}(\pi_j) G_{\alpha_i\alpha_j}(q_i,q_j,x_0) \nonumber \\
&\times \tilde\tau_{m'n'}^{\alpha_j\alpha_j'}(E_{q_j}) X_{m'\alpha_j',n\alpha_k}^{jk}(q_j,q_k;E) \nonumber \\
& -\int dq_j q_j^2 \bar Z_{m\alpha_i,m'\alpha_j}^{ij}(q_i,q_j,E) \frac{2M_j}{q_{0j}^2-q_j^2+i0}
\tilde\tau_{m'n'}^{\alpha_j\alpha_j'}(E_{q_j}) X_{n'\alpha_j',n\alpha_k}^{jk}(q_j,q_k;E)\Bigg\} .
\label{eeq:sing16}
\end{align}
For partial waves that do not support any two-body bound state, only the propagator singularity is present. To proceed, we define $N_b$ as the number of partial waves that support at least one two-body bound state so that the coupled equations for the
transition amplitudes take the form 
\begin{align}
X_{m\alpha_i,n\alpha_k}^{ik}(q_i,q_k;E)& = Z_{m\alpha_i,n\alpha_k}^{ik}(q_i,q_k,E)+
\sum\limits_{j}\Bigg[ \sum\limits_{\alpha_j=1}^{N_{bj}}\sum\limits_{\alpha_j'=1}^{N_{bj}} \sum\limits_{m'n'} \bar\delta_{ij}\Bigg\{ \nonumber \\
& \int dp_i  p_i\;   \frac{1}{\beta q_i}  h_m^{\alpha_i}(p_i)
 \frac{2\mu_i}{p_{0i}^2(q_i)-p_i^2+i0} \int\limits_{q_j=|p_i-\beta q_i|}^{q_j=p_i+\beta q_i} dq_j q_{j} \frac{1}{\frac{{p_j}^2}{2\mu_j}-\epsilon_{\alpha_j}}
h_{m'}^{\alpha_j}(\pi_j) G_{\alpha_i\alpha_j}(q_i,q_j,x_0) \nonumber \\
&\times \tilde\tau_{m'n'}^{\alpha_j\alpha_j'}(E_{q_j}) X_{m'\alpha_j',n\alpha_k}^{jk}(q_j,q_k;E) \nonumber \\
& -\int dq_j q_j^2 \bar Z_{m\alpha_i,n\alpha_j}^{ij}(q_i,q_j,E) \frac{2M_j}{q_{0j}^2-q_j^2+i0}
\tilde\tau_{m'n'}^{\alpha_j\alpha_j'}(E_{q_j}) X_{n'\alpha_j',n\alpha_k}^{jk}(q_j,q_k;E)\Bigg\}\cr
&+ \sum\limits_{\alpha_j>N_{bj}}\sum\limits_{\alpha_j'>N_{bj}} \sum\limits_{m'n'}\int dp_i  p_i\;   \frac{1}{\beta q_i}  h_m^{\alpha_i}(p_i)
 \frac{2\mu_i}{p_{0i}^2(q_i)-p_i^2+i0} \int\limits_{q_j=|p_i-\beta q_i|}^{q_j=p_i+\beta q_i} dq_j q_{j} 
h_{m'}^{\alpha_j}(\pi_j) G_{\alpha_i\alpha_j}(q_i,q_j,x_0)
\nonumber \\
&\times \tau_{m'n'}^{\alpha_j\alpha_j'}(E_{q_j}) X_{n'\alpha_j',n\alpha_k}^{jk}(q_j,q_k;E)\Bigg].
\label{eeq:sing17}
\end{align}   
At this point all the singularities of the kernel have been reduced to simple poles. Since
none of them are end point singularities, they can be removed using standard subtraction
techniques. The resulting integral equations with a regularized kernel can be readily solved using iterative Lanzcos-type techniques~\cite{Saad:2003}.

\section{The S-curve for Particles with Different Masses}

\label{scurve}
In a breakup configuration of three particles, the energy can be continuously 
distributed between the  relative motion of the fragments. Conservation of energy and
total momentum impose additional  constraints on
which breakup configurations are accessible. The kinematically allowed configurations
can be described by a so-called S-curve. 
In the following, the analytical form of this S-curve for  
three particles with different masses is derived in the laboratory frame. This is an
extension of the derivation presented in Ref.~\cite{Garrido:2014tma} where particles with
identical masses were considered.

First, the phase space $\Phi$ is given in Jacobi coordinates $\{\vec{p},\vec{q}\}$ as
\begin{equation}
\label{eq:jacobiphase}
\Phi=\int d\vec{p} \  d\vec{q} \ \delta (E-E_p-E_q),
\end{equation}
where the $\delta$-function is required by energy conservation. The three-body
energy in the c.m. frame is given by  $E$ and  
\begin{equation}
\int d\vec{p} \  d\vec{q} =  \int d\hat{p} \ d\hat{q} \  dE_p \ dE_q \  pq \mu_p\mu_q
,
\end{equation}
with $\mu_p$ and $\mu_q$ being the reduced masses. After integrating over $E_p$,
the phase space factor, Eq.~(\ref{eq:jacobiphase}),  takes the final form,
\begin{equation}
\Phi=\int d\hat{p} \ d\hat{q} \ dE_q \  pq\mu_p\mu_q.
\end{equation}
In the laboratory frame  the momentum of the target is zero before the collision.
We now write the phase-space factor $\Phi$  using the momenta $\vec{k}_i$,
$\vec{k}_j$, and $\vec{k}_k$ of the particles after the breakup. 
Imposing energy and momentum conservation $\Phi$  takes the form 
\begin{equation}
\Phi= \int d\vec{k}_i \ d\vec{k}_j \ d\vec{k}_k \  \delta \left(E_{lab}-\frac{k_i^2}{2m_i}
-\frac{k_j^2}{2m_j}-\frac{k_k^2}{2m_k}\right) \ 
\delta^3(\vec{P}_{lab}-\vec{k}_i-\vec{k}_j-\vec{k}_k), 
\end{equation}
where $\vec{P}_{lab}$ is the momentum of the incoming particle in the laboratory 
frame and $E_{lab}$ is the total three body energy in the laboratory frame. 
For the deuteron induced reaction the relation 
\begin{equation}
\label{eq:elab}
E_{lab}= E_{in}^{lab} + E_d = \frac{{P}_{lab}^2}{2m_d} + E_d
\end{equation}
holds, 
where $E_{in}^{lab}$ is the energy of the projectile in the laboratory frame, 
$m_d$ and $E_d$ are the mass and binding energy of the dimer, respectively. 

Making use of the momentum conserving $\delta$-function, one can integrate  over 
$\vec{k}_j$ which leads to 
\begin{equation}
\label{eq:phi_lab}
\Phi= \int d\vec{k}_i d\vec{k}_k \delta(f(\vec{k}_i, \vec{k}_k)), 
\end{equation}
where 
\begin{equation}
f(\vec{k}_i, \vec{k}_k) = E_{lab} - \frac{k_i^2}{2m_i} - \frac{k_k^2}{2m_k} - 
\frac{(\vec{P}_{lab}-\vec{k}_i-\vec{k}_k)^2}{2m_j}.
\end{equation}
Using Eq.~(\ref{eq:elab}) $f(\vec{k}_i, \vec{k}_k)$  can be written as 
\begin{equation}
f(\vec{k}_i, \vec{k}_k) = E_d + E_{in}^{lab}- \frac{P_{lab}^2}{2m_j}-
\frac{(m_i+m_j)k_i^2}{2m_im_j}-\frac{(m_j+m_k)k_k^2}{2m_jm_k}
+ \frac{P_{lab}k_i\omega_i+ P_{lab}k_k\omega_k- k_ik_k\omega}{m_j}, 
\end{equation}
with 
\begin{align} 
\omega_i & =  \frac{\vec{P}_{lab}\cdot \vec{k}_i}{P_{lab}k_i} = \cos \theta_{i},  \\ 
\omega_k & =  \frac{\vec{P}_{lab}\cdot \vec{k}_k}{P_{lab}k_k} = \cos \theta_{k}, \\
\omega &=  \frac{\vec{k}_i\cdot \vec{k}_k}{k_ik_k} = \sin\theta_i\sin\theta_k\cos\Delta\varphi + \cos \theta_{i}\cos \theta_{k}, 
\end{align}
where $\theta_i$ and $\theta_k$ are the polar angles associated the direction of 
$\vec{k}_i$ and $\vec{k}_k$  when assuming $\vec{P}_{lab}$ as $z-$axis. 
The quantity  $\Delta\varphi$ is the difference between the corresponding azimuthal angles. 

By using the well-known property of the $\delta$ function, 
$\delta[g(x)]=\sum_i\frac{1}{|g'(x_i)|}\delta(x-x_i)$ , it is easy to 
rewrite $f(\vec{k}_i, \vec{k}_k)$ in  a more convenient form 
and integrate Eq.~(\ref{eq:phi_lab}) over $k_i$, which leads to 
\begin{equation}
\label{eq:inter_ki}
\Phi= \int d\hat{k}_i \  d\hat{k}_k \  dk_k \  \frac{m_im_j k_i^2k_k^2}{|k_i(m_i+m_j)-m_iP_{lab}\omega_i+m_ik_k\omega|}. 
\end{equation} 
For each value of $k_k$ and the direction $\hat{k}_i$ and $\hat{k}_k$, $k_i$ 
is obtained from the solution of 
\begin{equation}
\label{eq:general_ellipse}
E_d + E_{in}^{lab}- \frac{P_{lab}^2}{2m_j} - \frac{(m_i+m_j)k_i^2}{2m_im_j}
-\frac{(m_j+m_k)k_k^2}{2m_jm_k} + \frac{P_{lab}k_i\omega_i + P_{lab}k_k\omega_k-k_ik_k\omega}{m_j}=0. 
\end{equation}
\noindent
In the same way, it is possible to integrate Eq~(\ref{eq:phi_lab}) over $k_k$ instead of $k_i$,
and obtain
\begin{equation}
\label{eq:inter_kk}
\Phi=\int d\hat{k}_i \  d\hat{k}_k \  dk_i \  \frac{m_jm_kk_i^2k_k^2}{|k_k(m_k+m_j)-m_kP_{lab}\omega_k+m_kk_i\omega|}. 
\end{equation}
From Eqs.~(\ref{eq:inter_ki}) and (\ref{eq:inter_kk}), one gets 
\begin{equation}
 \frac{m_i dk_k}{|k_i(m_i+m_j)-m_iP_{lab}\omega_i+m_ik_k\omega|} =  \frac{m_k dk_i}{|k_k(m_k+m_j)-m_kP_{lab}\omega_k+m_kk_i\omega|}. 
\end{equation}
Since the energy $E_i$ of particle $i$ is given by $E_i=\frac{k_i^2}{2m_i}$, we have $dE_i=\frac{k_i}{m_i}dk_i$,
and the expression above can be rewritten as 
\begin{equation}
 \frac{k_i dE_k}{|k_i(m_i+m_j)-m_iP_{lab}\omega_i+m_ik_k\omega|} =  \frac{k_k dE_i}{|k_k(m_k+m_j)-m_kP_{lab}\omega_k+m_kk_i\omega|}. 
\end{equation}
We now define $\mathcal{S}$ as
\begin{equation}
d\mathcal{S}= \sqrt{(dE_i)^2 + (dE_k)^2},
\end{equation}
which leads to the expression for phase space in laboratory frame 
\begin{equation}
\Phi=\int   d\hat{k}_i \  d\hat{k}_k \  d\mathcal{S} \ K_s, 
\end{equation}
with 
\begin{equation}
K_s = \frac{m_im_jm_kk_i^2k_k^2}{\sqrt{k_i^2|k_k(mk+mj)-m_kP_{lab}\omega_k+m_kk_i\omega|^2
+ k_k^2 |k_i(m_i+m_j)-m_iP_{lab}\omega_i+m_ik_k\omega|^2 }}.
\end{equation}

In this work the $d$-$\alpha$ breakup reaction is characterized 
by the incident energy of the projectile in the laboratory frame $E^{lab}_{in}$,
the polar angles $\theta_i$ and $\theta_k$ of the  particles observed 
after the breakup, and the difference between the corresponding two 
azimuthal angles $\Delta \varphi$. 
To obtain the cross section as a function of $\mathcal{S}$, one needs to determine 
the values of $k_i$ and $k_k$ which are related through Eq.~(\ref{eq:general_ellipse}). 
After some algebra the above equation can be cast into the form of an  ellipse, whose
characteristic  equation 
is given by
\begin{equation}
\label{eq:ellipse}
\frac{x^2}{a^2} + \frac{y^2}{b^2} = 1 
\end{equation}
with
\begin{equation}
\left[\!
    \begin{array}{c}
      x \\
      y
    \end{array}
  \!\right]=
  \left[\!
    \begin{array}{cc}
      \cos\tilde{\theta} & -\sin\tilde{\theta} \\
      \sin\tilde{\theta} & \cos\tilde{\theta}
    \end{array}
  \!\right] 
  \left[\!
    \begin{array}{c}
      k_i-k_i^c \\
      k_k-k_k^c
    \end{array}
  \!\right].
\end{equation}
Here $\tilde{\theta}$ is the angle of rotation, $k_i^c$ and $k_k^c$ are the center points 
for the $k_i$ and $k_k$ axes. These values are given by 
\begin{equation}
\tilde{\theta}=\frac{1}{2}\arctan(\frac{2m_im_k\omega}{m_j(m_i-m_k)}),
\end{equation}
\begin{equation}
k_i^c=\frac{ P_{lab}m_i(\omega_k\omega m_k-\omega_i(m_j+m_k)) }{ m_im_k\omega^2-(m_i+m_j)(m_j+m_k) },
\end{equation}
\begin{equation}
k_k^c=\frac{ P_{lab}m_k(\omega_i\omega m_i-\omega_k(m_j+m_i)) }{ m_im_k\omega^2-(m_i+m_j)(m_j+m_k) } ,
\end{equation}
and the axes of the ellipse $a$ and $b$ are given by 
\begin{equation}
a=\sqrt{ \frac{ \frac{m_j+m_i}{2m_i}k_i^{c2} + \frac{m_j+m_k}{2m_k}k_k^{c2}+k_i^ck_k^c\omega + m_jE_d - \frac{P_{lab}^2m_i}{2m_d} }{\frac{m_j(m_i+m_k)+2m_im_k}{4m_im_k}-\frac{\omega}{2\sin(2\tilde{\theta})}} },
\end{equation}

\begin{equation}
b=\sqrt{ \frac{ \frac{m_j+m_i}{2m_i}k_i^{c2} + \frac{m_j+m_k}{2m_k}k_k^{c2}+k_i^ck_k^c\omega + m_jE_d - \frac{P_{lab}^2m_i}{2m_d} }{\frac{m_j(m_i+m_k)+2m_im_k}{4m_im_k}+\frac{\omega}{2\sin(2\tilde{\theta})}} }.
\end{equation}
By using the parametric form for an ellipse, $k_i$ and $k_k$ can be written as 
\begin{equation}
k_i= a \cos t \cos\tilde{\theta}  + b\sin t\sin\tilde{\theta}  + k_i^c,
\end{equation}
\begin{equation}
k_k=- a \cos t \sin\tilde{\theta}  + b\sin t\cos\tilde{\theta}  + k_k^c, 
\end{equation}
where $t$ is the angular parameter. 
Making use of the two equations above, we can write the arc length $\mathcal{S}$
as 
\begin{equation}
d\mathcal{S} = \sqrt{\frac{k_i^2}{m_i^2}(-a\sin t \cos\tilde{\theta} + b\cos t \sin\tilde{\theta})^2 
+ \frac{k_k^2}{m_k^2} (a\sin t \sin\tilde{\theta}+ b\cos t \cos\tilde{\theta})^2} dt.
\end{equation}
Therefore, the arc length $\mathcal{S}$ can be obtained as a function of $t$
from the expression
\begin{equation}
\mathcal{S}(\tilde{t}) = \int_{t_0}^{\tilde{t}}\sqrt{\frac{k_i^2}{m_i^2}(-a\sin t \cos\tilde{\theta} + b\cos t \sin\tilde{\theta})^2 
+ \frac{k_k^2}{m_k^2} (a\sin t \sin\tilde{\theta}+ b\cos t \cos\tilde{\theta})^2} dt.
\end{equation}

\begin{acknowledgments}
This work was performed in part under the
auspices of the National Science Foundation under contract NSF-PHY-1520972
with Ohio University and NSF-PHY-1520929 with Michigan State University, 
of the U.~S.  Department of Energy under contract
No. DE-FG02-93ER40756 with Ohio University, and of DFG and NSFC through funds provided to the
Sino-German CRC 110 ``Symmetries and the Emergence of Structure in QCD" (NSFC
Grant No.~11621131001, DFG Grant No.~TRR110), and by the 
ExtreMe Matter Institute EMMI
at the GSI Helmholtzzentrum f\"ur Schwerionenphysik, Darmstadt, Germany.
This research used resources of the National Energy Research Scientific 
Computing Center (NERSC), a U.S. Department of Energy Office of Science 
User Facility operated under Contract No. DE-AC02-05CH11231.
This work was supported in part by Michigan State University through computational resources provided by the Institute for Cyber-Enabled Research. The work of A.Deltuva was supported by the Alexander von Humboldt-Foundation under
Grant No. LTU-1185721-HFST-E
\end{acknowledgments}

\bibliography{d_alpha,reactions}

\begin{thebibliography}{29}%
\makeatletter
\providecommand \@ifxundefined [1]{%
 \@ifx{#1\undefined}
}%
\providecommand \@ifnum [1]{%
 \ifnum #1\expandafter \@firstoftwo
 \else \expandafter \@secondoftwo
 \fi
}%
\providecommand \@ifx [1]{%
 \ifx #1\expandafter \@firstoftwo
 \else \expandafter \@secondoftwo
 \fi
}%
\providecommand \natexlab [1]{#1}%
\providecommand \enquote  [1]{``#1''}%
\providecommand \bibnamefont  [1]{#1}%
\providecommand \bibfnamefont [1]{#1}%
\providecommand \citenamefont [1]{#1}%
\providecommand \href@noop [0]{\@secondoftwo}%
\providecommand \href [0]{\begingroup \@sanitize@url \@href}%
\providecommand \@href[1]{\@@startlink{#1}\@@href}%
\providecommand \@@href[1]{\endgroup#1\@@endlink}%
\providecommand \@sanitize@url [0]{\catcode `\\12\catcode `\$12\catcode
  `\&12\catcode `\#12\catcode `\^12\catcode `\_12\catcode `\%12\relax}%
\providecommand \@@startlink[1]{}%
\providecommand \@@endlink[0]{}%
\providecommand \url  [0]{\begingroup\@sanitize@url \@url }%
\providecommand \@url [1]{\endgroup\@href {#1}{\urlprefix }}%
\providecommand \urlprefix  [0]{URL }%
\providecommand \Eprint [0]{\href }%
\providecommand \doibase [0]{http://dx.doi.org/}%
\providecommand \selectlanguage [0]{\@gobble}%
\providecommand \bibinfo  [0]{\@secondoftwo}%
\providecommand \bibfield  [0]{\@secondoftwo}%
\providecommand \translation [1]{[#1]}%
\providecommand \BibitemOpen [0]{}%
\providecommand \bibitemStop [0]{}%
\providecommand \bibitemNoStop [0]{.\EOS\space}%
\providecommand \EOS [0]{\spacefactor3000\relax}%
\providecommand \BibitemShut  [1]{\csname bibitem#1\endcsname}%
\let\auto@bib@innerbib\@empty
\bibitem [{\citenamefont {Atar}\ \emph {et~al.}(2018)\citenamefont {Atar} \emph
  {et~al.}}]{Atar:2018dhg}%
  \BibitemOpen
  \bibfield  {author} {\bibinfo {author} {\bibfnamefont {L.}~\bibnamefont
  {Atar}} \emph {et~al.},\ }\href {\doibase 10.1103/PhysRevLett.120.052501}
  {\bibfield  {journal} {\bibinfo  {journal} {Phys. Rev. Lett.}\ }\textbf
  {\bibinfo {volume} {120}},\ \bibinfo {pages} {052501} (\bibinfo {year}
  {2018})}\BibitemShut {NoStop}%
\bibitem [{\citenamefont {Shanley}(1969)}]{Shanley:1969zza}%
  \BibitemOpen
  \bibfield  {author} {\bibinfo {author} {\bibfnamefont {P.~E.}\ \bibnamefont
  {Shanley}},\ }\href {\doibase 10.1103/PhysRev.187.1328} {\bibfield  {journal}
  {\bibinfo  {journal} {Phys. Rev.}\ }\textbf {\bibinfo {volume} {187}},\
  \bibinfo {pages} {1328} (\bibinfo {year} {1969})}\BibitemShut {NoStop}%
\bibitem [{\citenamefont {Miyagawa}\ \emph {et~al.}(1985)\citenamefont
  {Miyagawa}, \citenamefont {Koike}, \citenamefont {Ueda}, \citenamefont
  {Sawada},\ and\ \citenamefont {Takagi}}]{Miyagawa:1985js}%
  \BibitemOpen
  \bibfield  {author} {\bibinfo {author} {\bibfnamefont {K.}~\bibnamefont
  {Miyagawa}}, \bibinfo {author} {\bibfnamefont {Y.}~\bibnamefont {Koike}},
  \bibinfo {author} {\bibfnamefont {T.}~\bibnamefont {Ueda}}, \bibinfo {author}
  {\bibfnamefont {T.}~\bibnamefont {Sawada}}, \ and\ \bibinfo {author}
  {\bibfnamefont {S.}~\bibnamefont {Takagi}},\ }\href {\doibase
  10.1143/PTP.74.1264} {\bibfield  {journal} {\bibinfo  {journal} {Prog. Theor.
  Phys.}\ }\textbf {\bibinfo {volume} {74}},\ \bibinfo {pages} {1264} (\bibinfo
  {year} {1985})}\BibitemShut {NoStop}%
\bibitem [{\citenamefont {Alt}\ \emph {et~al.}(1967{\natexlab{a}})\citenamefont
  {Alt}, \citenamefont {Grassberger},\ and\ \citenamefont {Sandhas}}]{ags1967}%
  \BibitemOpen
  \bibfield  {author} {\bibinfo {author} {\bibfnamefont {E.}~\bibnamefont
  {Alt}}, \bibinfo {author} {\bibfnamefont {P.}~\bibnamefont {Grassberger}}, \
  and\ \bibinfo {author} {\bibfnamefont {W.}~\bibnamefont {Sandhas}},\ }\href
  {\doibase https://doi.org/10.1016/0550-3213(67)90016-8} {\bibfield  {journal}
  {\bibinfo  {journal} {Nuclear Physics B}\ }\textbf {\bibinfo {volume} {2}},\
  \bibinfo {pages} {167 } (\bibinfo {year} {1967}{\natexlab{a}})}\BibitemShut
  {NoStop}%
\bibitem [{\citenamefont {Deltuva}(2006)}]{deltuva:06b}%
  \BibitemOpen
  \bibfield  {author} {\bibinfo {author} {\bibfnamefont {A.}~\bibnamefont
  {Deltuva}},\ }\href@noop {} {\bibfield  {journal} {\bibinfo  {journal}
  {Phys.~Rev.~C}\ }\textbf {\bibinfo {volume} {74}},\ \bibinfo {pages} {064001}
  (\bibinfo {year} {2006})}\BibitemShut {NoStop}%
\bibitem [{\citenamefont {Deltuva}\ and\ \citenamefont
  {Fonseca}(2009)}]{deltuva2009a}%
  \BibitemOpen
  \bibfield  {author} {\bibinfo {author} {\bibfnamefont {A.}~\bibnamefont
  {Deltuva}}\ and\ \bibinfo {author} {\bibfnamefont {A.~C.}\ \bibnamefont
  {Fonseca}},\ }\href {\doibase 10.1103/PhysRevC.79.014606} {\bibfield
  {journal} {\bibinfo  {journal} {Phys. Rev. C}\ }\textbf {\bibinfo {volume}
  {79}},\ \bibinfo {pages} {014606} (\bibinfo {year} {2009})}\BibitemShut
  {NoStop}%
\bibitem [{\citenamefont {Deltuva}\ \emph {et~al.}(2017)\citenamefont
  {Deltuva}, \citenamefont {Jur\v{c}iukonis},\ and\ \citenamefont
  {Norvai\v{s}as}}]{deltuva:17b}%
  \BibitemOpen
  \bibfield  {author} {\bibinfo {author} {\bibfnamefont {A.}~\bibnamefont
  {Deltuva}}, \bibinfo {author} {\bibfnamefont {D.}~\bibnamefont
  {Jur\v{c}iukonis}}, \ and\ \bibinfo {author} {\bibfnamefont {E.}~\bibnamefont
  {Norvai\v{s}as}},\ }\href@noop {} {\bibfield  {journal} {\bibinfo  {journal}
  {Phys. Lett. B}\ }\textbf {\bibinfo {volume} {769}},\ \bibinfo {pages} {202}
  (\bibinfo {year} {2017})}\BibitemShut {NoStop}%
\bibitem [{\citenamefont {Mukhamedzhanov}\ \emph {et~al.}(2012)\citenamefont
  {Mukhamedzhanov}, \citenamefont {Eremenko},\ and\ \citenamefont
  {Sattarov}}]{akram2012}%
  \BibitemOpen
  \bibfield  {author} {\bibinfo {author} {\bibfnamefont {A.~M.}\ \bibnamefont
  {Mukhamedzhanov}}, \bibinfo {author} {\bibfnamefont {V.}~\bibnamefont
  {Eremenko}}, \ and\ \bibinfo {author} {\bibfnamefont {A.~I.}\ \bibnamefont
  {Sattarov}},\ }\href {\doibase 10.1103/PhysRevC.86.034001} {\bibfield
  {journal} {\bibinfo  {journal} {Phys. Rev. C}\ }\textbf {\bibinfo {volume}
  {86}},\ \bibinfo {pages} {034001} (\bibinfo {year} {2012})}\BibitemShut
  {NoStop}%
\bibitem [{\citenamefont {Lovelace}(1964{\natexlab{a}})}]{lovelace1964}%
  \BibitemOpen
  \bibfield  {author} {\bibinfo {author} {\bibfnamefont {C.}~\bibnamefont
  {Lovelace}},\ }\href {\doibase 10.1103/PhysRev.135.B1225} {\bibfield
  {journal} {\bibinfo  {journal} {Phys. Rev.}\ }\textbf {\bibinfo {volume}
  {135}},\ \bibinfo {pages} {B1225} (\bibinfo {year}
  {1964}{\natexlab{a}})}\BibitemShut {NoStop}%
\bibitem [{\citenamefont {Hlophe}\ \emph
  {et~al.}(2013{\natexlab{a}})\citenamefont {Hlophe}, \citenamefont {Elster},
  \citenamefont {Johnson}, \citenamefont {Upadhyay}, \citenamefont {Nunes},
  \citenamefont {Arbanas}, \citenamefont {Eremenko}, \citenamefont {Escher},\
  and\ \citenamefont {Thompson}}]{hlophe2013}%
  \BibitemOpen
  \bibfield  {author} {\bibinfo {author} {\bibfnamefont {L.}~\bibnamefont
  {Hlophe}}, \bibinfo {author} {\bibfnamefont {C.}~\bibnamefont {Elster}},
  \bibinfo {author} {\bibfnamefont {R.~C.}\ \bibnamefont {Johnson}}, \bibinfo
  {author} {\bibfnamefont {N.~J.}\ \bibnamefont {Upadhyay}}, \bibinfo {author}
  {\bibfnamefont {F.~M.}\ \bibnamefont {Nunes}}, \bibinfo {author}
  {\bibfnamefont {G.}~\bibnamefont {Arbanas}}, \bibinfo {author} {\bibfnamefont
  {V.}~\bibnamefont {Eremenko}}, \bibinfo {author} {\bibfnamefont {J.~E.}\
  \bibnamefont {Escher}}, \ and\ \bibinfo {author} {\bibfnamefont {I.~J.}\
  \bibnamefont {Thompson}} (\bibinfo {collaboration} {TORUS Collaboration}),\
  }\href {\doibase 10.1103/PhysRevC.88.064608} {\bibfield  {journal} {\bibinfo
  {journal} {Phys. Rev. C}\ }\textbf {\bibinfo {volume} {88}},\ \bibinfo
  {pages} {064608} (\bibinfo {year} {2013}{\natexlab{a}})}\BibitemShut
  {NoStop}%
\bibitem [{\citenamefont {Hlophe}\ \emph {et~al.}(2014)\citenamefont {Hlophe},
  \citenamefont {Eremenko}, \citenamefont {Elster}, \citenamefont {Nunes},
  \citenamefont {Arbanas}, \citenamefont {Escher},\ and\ \citenamefont
  {Thompson}}]{hlophe2014}%
  \BibitemOpen
  \bibfield  {author} {\bibinfo {author} {\bibfnamefont {L.}~\bibnamefont
  {Hlophe}}, \bibinfo {author} {\bibfnamefont {V.}~\bibnamefont {Eremenko}},
  \bibinfo {author} {\bibfnamefont {C.}~\bibnamefont {Elster}}, \bibinfo
  {author} {\bibfnamefont {F.~M.}\ \bibnamefont {Nunes}}, \bibinfo {author}
  {\bibfnamefont {G.}~\bibnamefont {Arbanas}}, \bibinfo {author} {\bibfnamefont
  {J.~E.}\ \bibnamefont {Escher}}, \ and\ \bibinfo {author} {\bibfnamefont
  {I.~J.}\ \bibnamefont {Thompson}} (\bibinfo {collaboration} {TORUS
  Collaboration}),\ }\href {\doibase 10.1103/PhysRevC.90.061602} {\bibfield
  {journal} {\bibinfo  {journal} {Phys. Rev. C}\ }\textbf {\bibinfo {volume}
  {90}},\ \bibinfo {pages} {061602} (\bibinfo {year} {2014})}\BibitemShut
  {NoStop}%
\bibitem [{\citenamefont {Hlophe}\ \emph
  {et~al.}(2017{\natexlab{a}})\citenamefont {Hlophe}, \citenamefont {Lei},
  \citenamefont {Elster}, \citenamefont {Nogga},\ and\ \citenamefont
  {Nunes}}]{hlophe2017}%
  \BibitemOpen
  \bibfield  {author} {\bibinfo {author} {\bibfnamefont {L.}~\bibnamefont
  {Hlophe}}, \bibinfo {author} {\bibfnamefont {J.}~\bibnamefont {Lei}},
  \bibinfo {author} {\bibfnamefont {C.}~\bibnamefont {Elster}}, \bibinfo
  {author} {\bibfnamefont {A.}~\bibnamefont {Nogga}}, \ and\ \bibinfo {author}
  {\bibfnamefont {F.~M.}\ \bibnamefont {Nunes}},\ }\href {\doibase
  10.1103/PhysRevC.96.064003} {\bibfield  {journal} {\bibinfo  {journal} {Phys.
  Rev. C}\ }\textbf {\bibinfo {volume} {96}},\ \bibinfo {pages} {064003}
  (\bibinfo {year} {2017}{\natexlab{a}})}\BibitemShut {NoStop}%
\bibitem [{\citenamefont {Witala}\ and\ \citenamefont
  {Gl\"ockle}(2008)}]{Witala:2008my}%
  \BibitemOpen
  \bibfield  {author} {\bibinfo {author} {\bibfnamefont {H.}~\bibnamefont
  {Witala}}\ and\ \bibinfo {author} {\bibfnamefont {W.}~\bibnamefont
  {Gl\"ockle}},\ }\href {\doibase 10.1140/epja/i2008-10610-x} {\bibfield
  {journal} {\bibinfo  {journal} {Eur. Phys. J.}\ }\textbf {\bibinfo {volume}
  {A37}},\ \bibinfo {pages} {87} (\bibinfo {year} {2008})},\ \Eprint
  {http://arxiv.org/abs/0806.2757} {arXiv:0806.2757 [nucl-th]} \BibitemShut
  {NoStop}%
\bibitem [{\citenamefont {Elster}\ \emph {et~al.}(2009)\citenamefont {Elster},
  \citenamefont {Gl\"ockle},\ and\ \citenamefont {Witala}}]{Elster:2008hn}%
  \BibitemOpen
  \bibfield  {author} {\bibinfo {author} {\bibfnamefont {C.}~\bibnamefont
  {Elster}}, \bibinfo {author} {\bibfnamefont {W.}~\bibnamefont {Gl\"ockle}}, \
  and\ \bibinfo {author} {\bibfnamefont {H.}~\bibnamefont {Witala}},\ }\href
  {\doibase 10.1007/s00601-008-0003-6} {\bibfield  {journal} {\bibinfo
  {journal} {Few Body Syst.}\ }\textbf {\bibinfo {volume} {45}},\ \bibinfo
  {pages} {1} (\bibinfo {year} {2009})},\ \Eprint
  {http://arxiv.org/abs/0807.1421} {arXiv:0807.1421 [nucl-th]} \BibitemShut
  {NoStop}%
\bibitem [{\citenamefont {Gl\"ockle}(1983)}]{Gloecklebook}%
  \BibitemOpen
  \bibfield  {author} {\bibinfo {author} {\bibfnamefont {W.}~\bibnamefont
  {Gl\"ockle}},\ }\href@noop {} {\emph {\bibinfo {title} {The Quantum
  Mechanical Few-Body Problem}}},\ Texts and Monographs in Physics\ (\bibinfo
  {publisher} {Springer Verlag},\ \bibinfo {year} {1983})\BibitemShut {NoStop}%
\bibitem [{\citenamefont {Hlophe}\ and\ \citenamefont
  {Elster}(2017)}]{hlophe2017est}%
  \BibitemOpen
  \bibfield  {author} {\bibinfo {author} {\bibfnamefont {L.}~\bibnamefont
  {Hlophe}}\ and\ \bibinfo {author} {\bibfnamefont {C.}~\bibnamefont
  {Elster}},\ }\href {\doibase 10.1103/PhysRevC.95.054617} {\bibfield
  {journal} {\bibinfo  {journal} {Phys. Rev. C}\ }\textbf {\bibinfo {volume}
  {95}},\ \bibinfo {pages} {054617} (\bibinfo {year} {2017})}\BibitemShut
  {NoStop}%
\bibitem [{\citenamefont {Nemoto}\ \emph {et~al.}(1998)\citenamefont {Nemoto},
  \citenamefont {Chmielewski}, \citenamefont {Schellingerhout}, \citenamefont
  {Haidenbauer}, \citenamefont {Oryu},\ and\ \citenamefont
  {Sauer}}]{nemoto1998}%
  \BibitemOpen
  \bibfield  {author} {\bibinfo {author} {\bibfnamefont {S.}~\bibnamefont
  {Nemoto}}, \bibinfo {author} {\bibfnamefont {K.}~\bibnamefont {Chmielewski}},
  \bibinfo {author} {\bibfnamefont {N.~W.}\ \bibnamefont {Schellingerhout}},
  \bibinfo {author} {\bibfnamefont {J.}~\bibnamefont {Haidenbauer}}, \bibinfo
  {author} {\bibfnamefont {S.}~\bibnamefont {Oryu}}, \ and\ \bibinfo {author}
  {\bibfnamefont {P.~U.}\ \bibnamefont {Sauer}},\ }\href {\doibase
  10.1007/s006010050087} {\bibfield  {journal} {\bibinfo  {journal} {Few-Body
  Systems}\ }\textbf {\bibinfo {volume} {24}},\ \bibinfo {pages} {213}
  (\bibinfo {year} {1998})}\BibitemShut {NoStop}%
\bibitem [{\citenamefont {Cornelius}\ \emph {et~al.}(1990)\citenamefont
  {Cornelius}, \citenamefont {Gl\"ockle}, \citenamefont {Haidenbauer},
  \citenamefont {Koike}, \citenamefont {Plessas},\ and\ \citenamefont
  {Witala}}]{Cornelius:1990zz}%
  \BibitemOpen
  \bibfield  {author} {\bibinfo {author} {\bibfnamefont {T.}~\bibnamefont
  {Cornelius}}, \bibinfo {author} {\bibfnamefont {W.}~\bibnamefont
  {Gl\"ockle}}, \bibinfo {author} {\bibfnamefont {J.}~\bibnamefont
  {Haidenbauer}}, \bibinfo {author} {\bibfnamefont {Y.}~\bibnamefont {Koike}},
  \bibinfo {author} {\bibfnamefont {W.}~\bibnamefont {Plessas}}, \ and\
  \bibinfo {author} {\bibfnamefont {H.}~\bibnamefont {Witala}},\ }\href
  {\doibase 10.1103/PhysRevC.41.2538} {\bibfield  {journal} {\bibinfo
  {journal} {Phys. Rev.}\ }\textbf {\bibinfo {volume} {C41}},\ \bibinfo {pages}
  {2538} (\bibinfo {year} {1990})}\BibitemShut {NoStop}%
\bibitem [{\citenamefont {Alt}\ \emph {et~al.}(1967{\natexlab{b}})\citenamefont
  {Alt}, \citenamefont {Grassberger},\ and\ \citenamefont {Sandhas}}]{AGS}%
  \BibitemOpen
  \bibfield  {author} {\bibinfo {author} {\bibfnamefont {E.~O.}\ \bibnamefont
  {Alt}}, \bibinfo {author} {\bibfnamefont {P.}~\bibnamefont {Grassberger}}, \
  and\ \bibinfo {author} {\bibfnamefont {W.}~\bibnamefont {Sandhas}},\ }\href
  {\doibase 10.1016/0550-3213(67)90016-8} {\bibfield  {journal} {\bibinfo
  {journal} {Nucl. Phys.}\ }\textbf {\bibinfo {volume} {B2}},\ \bibinfo {pages}
  {167} (\bibinfo {year} {1967}{\natexlab{b}})}\BibitemShut {NoStop}%
\bibitem [{\citenamefont {Ernst}\ \emph {et~al.}(1973)\citenamefont {Ernst},
  \citenamefont {Shakin},\ and\ \citenamefont {Thaler}}]{Ernst:1973zzb}%
  \BibitemOpen
  \bibfield  {author} {\bibinfo {author} {\bibfnamefont {D.~J.}\ \bibnamefont
  {Ernst}}, \bibinfo {author} {\bibfnamefont {C.~M.}\ \bibnamefont {Shakin}}, \
  and\ \bibinfo {author} {\bibfnamefont {R.~M.}\ \bibnamefont {Thaler}},\
  }\href {\doibase 10.1103/PhysRevC.8.46} {\bibfield  {journal} {\bibinfo
  {journal} {Phys.Rev.}\ }\textbf {\bibinfo {volume} {C8}},\ \bibinfo {pages}
  {46} (\bibinfo {year} {1973})}\BibitemShut {NoStop}%
\bibitem [{\citenamefont {Lovelace}(1964{\natexlab{b}})}]{Lovelace:1964}%
  \BibitemOpen
  \bibfield  {author} {\bibinfo {author} {\bibfnamefont {C.}~\bibnamefont
  {Lovelace}},\ }\href {\doibase 10.1103/PhysRev.135.B1225} {\bibfield
  {journal} {\bibinfo  {journal} {Phys.Rev.}\ }\textbf {\bibinfo {volume}
  {135}},\ \bibinfo {pages} {B1225} (\bibinfo {year}
  {1964}{\natexlab{b}})}\BibitemShut {NoStop}%
\bibitem [{\citenamefont {Saad}(2003)}]{Saad:2003}%
  \BibitemOpen
  \bibfield  {author} {\bibinfo {author} {\bibfnamefont {Y.}~\bibnamefont
  {Saad}},\ }\href@noop {} {\emph {\bibinfo {title} {{Iterative Methods for
  Sparse Linear Systems}}}}\ (\bibinfo  {publisher} {Society for Industrial and
  Applied Mathematics, SIAM},\ \bibinfo {year} {2003})\BibitemShut {NoStop}%
\bibitem [{\citenamefont {Hlophe}\ \emph
  {et~al.}(2017{\natexlab{b}})\citenamefont {Hlophe}, \citenamefont {Lei},
  \citenamefont {Elster}, \citenamefont {Nogga},\ and\ \citenamefont
  {Nunes}}]{Hlophe:2017bkd}%
  \BibitemOpen
  \bibfield  {author} {\bibinfo {author} {\bibfnamefont {L.}~\bibnamefont
  {Hlophe}}, \bibinfo {author} {\bibfnamefont {J.}~\bibnamefont {Lei}},
  \bibinfo {author} {\bibfnamefont {C.}~\bibnamefont {Elster}}, \bibinfo
  {author} {\bibfnamefont {A.}~\bibnamefont {Nogga}}, \ and\ \bibinfo {author}
  {\bibfnamefont {F.~M.}\ \bibnamefont {Nunes}},\ }\href {\doibase
  10.1103/PhysRevC.96.064003} {\bibfield  {journal} {\bibinfo  {journal} {Phys.
  Rev.}\ }\textbf {\bibinfo {volume} {C96}},\ \bibinfo {pages} {064003}
  (\bibinfo {year} {2017}{\natexlab{b}})}\BibitemShut {NoStop}%
\bibitem [{\citenamefont {Machleidt}(2001)}]{Machleidt:2000ge}%
  \BibitemOpen
  \bibfield  {author} {\bibinfo {author} {\bibfnamefont {R.}~\bibnamefont
  {Machleidt}},\ }\href {\doibase 10.1103/PhysRevC.63.024001} {\bibfield
  {journal} {\bibinfo  {journal} {Phys. Rev.}\ }\textbf {\bibinfo {volume}
  {C63}},\ \bibinfo {pages} {024001} (\bibinfo {year} {2001})}\BibitemShut
  {NoStop}%
\bibitem [{\citenamefont {Thompson}\ \emph {et~al.}(2000)\citenamefont
  {Thompson}, \citenamefont {Danilin}, \citenamefont {Efros}, \citenamefont
  {Vaagen}, \citenamefont {Bang},\ and\ \citenamefont {Zhukov}}]{Thompson2000}%
  \BibitemOpen
  \bibfield  {author} {\bibinfo {author} {\bibfnamefont {I.~J.}\ \bibnamefont
  {Thompson}}, \bibinfo {author} {\bibfnamefont {B.~V.}\ \bibnamefont
  {Danilin}}, \bibinfo {author} {\bibfnamefont {V.~D.}\ \bibnamefont {Efros}},
  \bibinfo {author} {\bibfnamefont {J.~S.}\ \bibnamefont {Vaagen}}, \bibinfo
  {author} {\bibfnamefont {J.~M.}\ \bibnamefont {Bang}}, \ and\ \bibinfo
  {author} {\bibfnamefont {M.~V.}\ \bibnamefont {Zhukov}},\ }\href {\doibase
  10.1103/PhysRevC.61.024318} {\bibfield  {journal} {\bibinfo  {journal} {Phys.
  Rev. C}\ }\textbf {\bibinfo {volume} {61}},\ \bibinfo {pages} {024318}
  (\bibinfo {year} {2000})}\BibitemShut {NoStop}%
\bibitem [{\citenamefont {Ernst}\ \emph {et~al.}(1974)\citenamefont {Ernst},
  \citenamefont {Shakin},\ and\ \citenamefont {Thaler}}]{Ernst:1974zza}%
  \BibitemOpen
  \bibfield  {author} {\bibinfo {author} {\bibfnamefont {D.~J.}\ \bibnamefont
  {Ernst}}, \bibinfo {author} {\bibfnamefont {C.~M.}\ \bibnamefont {Shakin}}, \
  and\ \bibinfo {author} {\bibfnamefont {R.~M.}\ \bibnamefont {Thaler}},\
  }\href {\doibase 10.1103/PhysRevC.9.1780} {\bibfield  {journal} {\bibinfo
  {journal} {Phys.Rev.}\ }\textbf {\bibinfo {volume} {C9}},\ \bibinfo {pages}
  {1780} (\bibinfo {year} {1974})}\BibitemShut {NoStop}%
\bibitem [{\citenamefont {Hlophe}\ \emph
  {et~al.}(2013{\natexlab{b}})\citenamefont {Hlophe} \emph
  {et~al.}}]{Hlophe:2013xca}%
  \BibitemOpen
  \bibfield  {author} {\bibinfo {author} {\bibfnamefont {L.}~\bibnamefont
  {Hlophe}} \emph {et~al.} (\bibinfo {collaboration} {The TORUS
  Collaboration}),\ }\href {\doibase 10.1103/PhysRevC.88.064608} {\bibfield
  {journal} {\bibinfo  {journal} {Phys.Rev.}\ }\textbf {\bibinfo {volume}
  {C88}},\ \bibinfo {pages} {064608} (\bibinfo {year}
  {2013}{\natexlab{b}})}\BibitemShut {NoStop}%
\bibitem [{\citenamefont {Garrido}\ \emph {et~al.}(2014)\citenamefont
  {Garrido}, \citenamefont {Kievsky},\ and\ \citenamefont
  {Viviani}}]{Garrido:2014tma}%
  \BibitemOpen
  \bibfield  {author} {\bibinfo {author} {\bibfnamefont {E.}~\bibnamefont
  {Garrido}}, \bibinfo {author} {\bibfnamefont {A.}~\bibnamefont {Kievsky}}, \
  and\ \bibinfo {author} {\bibfnamefont {M.}~\bibnamefont {Viviani}},\ }\href
  {\doibase 10.1103/PhysRevC.90.014607} {\bibfield  {journal} {\bibinfo
  {journal} {Phys. Rev.}\ }\textbf {\bibinfo {volume} {C90}},\ \bibinfo {pages}
  {014607} (\bibinfo {year} {2014})},\ \Eprint {http://arxiv.org/abs/1407.0172}
  {arXiv:1407.0172 [nucl-th]} \BibitemShut {NoStop}%
\bibitem [{\citenamefont {Bang}\ and\ \citenamefont
  {Gignoux}(1979)}]{Bang:1979ihm}%
  \BibitemOpen
  \bibfield  {author} {\bibinfo {author} {\bibfnamefont {J.}~\bibnamefont
  {Bang}}\ and\ \bibinfo {author} {\bibfnamefont {C.}~\bibnamefont {Gignoux}},\
  }\href {\doibase 10.1016/0375-9474(79)90571-2} {\bibfield  {journal}
  {\bibinfo  {journal} {Nucl. Phys.}\ }\textbf {\bibinfo {volume} {A313}},\
  \bibinfo {pages} {119} (\bibinfo {year} {1979})}\BibitemShut {NoStop}%
\end{thebibliography}%

\newpage

\newpage
\clearpage

\begin{table}[ht]
\centering
\begin{tabular}{ccccc}

\hline\hline
&&\\
{ label} & & rank &support energy $E_n$ [MeV] & support momenta $p_n$ [fm$^{-1}$]\\
&&\\
 \hline
&&\\
NN-EST3-1 &&3&$-20,\;-2,\;20$& $\;0.7\;0.2,\;0.5$\\
NN-EST4-1& &4&$-60,\;-20,10,\;50$ &$\;1.2,\;0.7,\;3.0,\;0.8$\\ 
NN-EST5-1&&5&$-80,\;-40,\;-5,\;10,\;50$ &$\;1.4,\;1.0,\;0.3,\;0.3,\;0.8$\\ 
%
NN-EST6-1&&6&$\;-100,\;-60,\;-2,\;10,\;35,\;50$ & $\;1.6,\;1.2,\;0.2,\;0.3,\;0.7,\;0.8$\\
%
NN-EST6-2 &&6&$-20,\;-20,\;-20,\;-3,\;-3,\;-3$ & $\;0.4,\;0.7,\;1.1,\;0.4,\;1.1,\;2.5$\\
NN-EST6-3 &&6&$-20,\;-20,\;-20,\;30,\;30\;,30$ & $\;0.4,\;0.7,\;1.1,\;0.4,\;1.1,\;2.5$\\
NN-EST7-1&&7&$-150,\;-50,\;-25,\;-2,\;10,\;35,\;75$ &$\;2.4,\;1.4,\;1.0,\;0.3,\;0.6,\;1.0,\;1.2$\\
NN-EST7-2&&7&$-20,-20,-20,-2,-3,-3, -3$ &$\;0.2,\;0.4,\;1.1,\;3.0,\;0.4,\;1.0,\;3.0 $\\
\hline

\end{tabular}
\caption{Separable expansion  of the CD-Bonn~\cite{Machleidt:2000ge} potential. The  labels and ranks 
are listed in the first and second column. The
corresponding support energies and momenta  are given in the third
and fourth columns.      
  }
      \label{table1}
\end{table}

\begin{table}[ht]
\centering
\begin{tabular}{ccccc}

\hline\hline
&&\\
{ label}  && rank &support energy $E_n$ [MeV] & support momenta $p_n$ [fm$^{-1}$]\\
&&\\
 \hline
&&\\
NA-EST3-1 &&3&$-50,\;10,\;15$& $\;1.4,\;0.6,\;0.7$\\
NA-EST4-1&&4&$-50,\;-5,\;10,\;35$ &$\;1.4,\;0.4,\;0.6,\;1.0$\\  
%
NA-EST5-1&&5&$-80,\;-40,\;-1,\;10,\;50$ &$\;1.8,\;1.2,\;0.2,\;0.6,\;1.2$\\ 
NA-EST6-1&& 6&$-140,\;-100,\;-60,\;-20,\;-20,\;40$ & $\;2.3,\;2.0,\;1.5,\;0.9,\;0.8,\;1.1$\\
NA-EST6-2&&6&$-25,-25,-25,\;-5,\;-5,\;-5$ &$\;0.4,\;1.1,\;2.0,\;0.4,\;1.1,\;2.0 $\\
NA-EST6-3&&6&$-25,-25,-25,\;35,\;35,\;35$ &$\;1.0,\;1.2,\;3.,\;1.0,\;2.0,\;3.0 $\\  
NA-EST7-1&&7&$-150,\;-50,\;-25,\;-2,\;10,\;35,\;75$ &$\;2.4,\;1.4,\;1.0,\;0.3,\;0.6,\;1.0,\;1.2$\\
NA-EST7-2&&7&$-20,-20,-20,-2,-3,-3, -3$ &$\;0.2,\;0.4,\;1.1,\;3.,\;0.4,\;1.0,\;3.0 $\\
  %
\hline

\end{tabular}
\caption{Separable expansion of the Bang potential~\cite{Bang:1979ihm}. The  labels and ranks 
are listed in the first and second column. The
corresponding support energies and momenta  are given in the third
and fourth columns.      
  }
      \label{table2}
\end{table}

\begin{table}[ht]
\centering
\begin{tabular}{cccccccc}
\hline\hline
&&\\
 &&{\bf potential}&&  {\bf cross section~[mb]} & &\\
 \hline
&np& ~~~~~~~~~~~n/p-$\alpha$&~~~~elastic&breakup&total\\
\hline
&NN-EST3-1&NA-EST3-1& 789.1 & 492.3 & 1281.4\\
&NN-EST4-1&NA-EST4-1& 782.8&  503.7 & 1286.5\\
&NN-EST5-1&NA-EST5-1& 780.9 & 502.4 & 1283.3 \\
&NN-EST6-1&NA-EST6-1& 780.9 & 505.7 & 1286.5  \\
&NN-EST7-1&NA-EST7-1& 781.0 & 503.9 & 1284.9 \\
%
&\\
&NN-EST6-2&NA-EST6-2& 780.9 & 504.2& 1285.1\\
&NN-EST6-3&NA-EST6-3& 781.0 & 504.2& 1285.2 \\
&NN-EST7-2&NA-EST7-2& 780.6 & 503.9& 1284.5 \\
%
&\\
&CD-Bonn & Bang& {\bf 781.8} & {\bf 505.8} & {\bf 1287.6}\\
%
%
\hline
\end{tabular}
\caption{The integrated cross sections for d+$\alpha$ scattering obtained with the
separable potentials of Tables~\ref{table1} and~\ref{table2} at 10~MeV incident deuteron
energy. The potentials for
the np and n/p-$\alpha$ reactions are given in the first and second columns, respectively.
The cross section for elastic scattering and deuteron breakup are listed in the
third and fourth columns while the total cross section is presented in the last column.  
and fourth columns. The exact results obtained using the original CD-Bonn and Bang interaction are displayed  in the last row.        
  }
      \label{table3}
\end{table}
\begin{table}[ht]
\centering
\begin{tabular}{cccccccc}
\hline\hline
&&\\
 &&{\bf potential}&&  {\bf cross section~[mb]} & &\\
 \hline
&np& ~~~~n/p-$\alpha$&~~~~elastic&breakup&total\\
\hline
&NN-EST3-1&NA-EST3-1& 814.6 & 540.8 & 1355.4 \\
&NN-EST4-1&NA-EST4-1& 812.1 &540.9 & 1353.0 \\
&NN-EST5-1&NA-EST5-1& 811.0 & 540.3 & 1351.5 \\
&NN-EST6-1&NA-EST6-1& 810.6 & 541.7  & 1352.3\\
&NN-EST7-1&NA-EST7-1& 808.2 & 543.2  & 1351.5 \\

%
&\\
&NN-EST6-2&NA-EST6-2& 810.5 & 541.7 & 1352.2 \\
&NN-EST6-3&NA-EST6-3& 810.2 & 541.6 & 1351.8 \\
&NN-EST7-2&NA-EST7-2& 810.4 & 541.6 & 1352.0\\
%

&\\
&CD-Bonn&Bang& {\bf 809.2}& {\bf 542.5} & {\bf 1351.6}\\
%
%
\hline
\end{tabular}
\caption{Same as Table~\ref{table3} but for 20~MeV incident deuteron
energy.     
  }
      \label{table4}
\end{table}

\newpage
\clearpage
\noindent

\end{document}